\documentclass[aps,prd,preprint,showkeys,nofootinbib]{revtex4-1}
\usepackage[utf8]{inputenc}
\usepackage{amsmath}
\usepackage{amsfonts}
\usepackage{amssymb}
\usepackage{hyperref}
\usepackage{color}
\usepackage{graphicx}
\usepackage{epstopdf}

\usepackage{verbatim}
\usepackage{amsmath}
\usepackage{amsfonts}
\usepackage{cancel}
\usepackage{braket}
\usepackage{hyperref}
\hypersetup{
    colorlinks=true,
    linkcolor=blue,
    urlcolor=cyan,
    citecolor=cyan,}
\newcommand{\s}{\hspace{0.01cm}}

\usepackage{fancyvrb}
\usepackage{subfig}
\numberwithin{equation}{section}

\begin{document}

\title{Out of equilibrium Chiral Vortical Effect in Holography}
\author{Sergio Morales Tejera}
\email{sergio.moralest@uam.es}
\author{Karl Landsteiner}
\email{karl.landsteiner@csic.es}

\affiliation{Instituto de F\'isica Te\'orica UAM/CSIC, c/Nicol\'as Cabrera 13-15, Universidad Aut\'onoma de Madrid, Cantoblanco, 28049 Madrid, Spain}
\preprint{IFT-UAM/CSIC-20-92}
\date{\today}

\begin{abstract}
We study the chiral vortical effect far from equilibrium in a strongly coupled holographic field theory. Rotation is represented as a perturbation via a gravito-magnetic field on top of a five-dimensional charged AdS Vaidya metric. We also introduce a momentum relaxation mechanism by linear scalar field backgrounds and study the CVE dynamics as function of the charges, temperature and momentum relaxation. The far from equilibrium behavior shows that the CVE builds up with a significant delay in time compared to the quasi instantaneous equilibration of the background metric. We also pay special attention to the effects of the gravitational contribution to the axial anomaly in the CVE of the axial current.
We develop an analytic estimate of this delay and also compute the quasi-normal modes near equilibrium which determine the late time ring down. 
\end{abstract}

\maketitle

\newpage
\tableofcontents

\newpage

\section{Introduction}
Anomaly induced transport phenomena such as the Chiral Magnetic (CME) and Chiral Vortical (CVE) Effects are extremely active areas of research (see  \cite{Kharzeev:2015znc, Landsteiner:2016led} for reviews). They play important roles in high energy as well as in condensed matter physics. In high energy physics the search for signatures of anomaly induced transport in heavy ion collisions is an on-going endeavor and has culminated in the iso-bar program at RHIC \cite{Kharzeev:2019zgg}. The CVE can also  be important in heavy ion collisions especially due to the proven presence of large vorticity \cite{STAR:2017ckg}.

 In condensed matter physics the CME is at the origin of the observed large enhancement of the longitudinal conductivity of Weyl- and Dirac semimetals subject to a magnetic field \cite{Li:2014bha}. The CVE has so far played a minor role in condensed matter physics.

While formally both CME and CVE can be derived in mathematical models of equilibrium quantum field theory, both are in a subtle way manifestations of out-of-equilibrium physics. For the CME it is now well established that it vanishes in strict equilibrium \cite{Gynther:2010ed, FranzVazifeh,Yamamoto:2015fxa}. Theoretically it is due to a non-trivial contribution coming from the vacuum of the filled Dirac sea giving rise to the so-called Bardeen counterterms \cite{Bardeen:1969md}. For the CVE the situation is slightly different. While it is possible to derive it in a way completely analogous to the CME it often relies formally on a thermal ensemble at fixed angular velocity. In the context of the CVE it has been noted very early on that an equilibrium thermal ensemble with fixed angular velocity is not possible in a relativistic context. The tangential velocity necessarily exceeds the speed of light at some finite distance from the axis of rotation \cite{Vilenkin:1980zv}. A useful tool to study the CVE is  restricting to infinitesimal angular velocity which implies linear response theory \footnote{An equivalent approach is to consider an ensemble in a finite cylinder of radius $R$ and compute the response at the axes of rotation in the limit of $T/R \rightarrow 0$.}. In this regime it has been established that a useful way of thinking about the CVE is not directly as a rotation but as a gravito-magnetic analogue of the CME \cite{Landsteiner:2011tg}. In this approach one studies a fluid at rest in an (infinitesimally small) gravitational field involving a mixed time-space component of the metric. This approach gives rise to the Kubo formula for the CVE and lies also at the heart of its study in holographic models \cite{Amado:2011zx,Erdmenger:2008rm, Banerjee:2008th}. 

The theoretical tools described above allow to compute the response of anomalous field theories to a magnetic field or rotation in  near (or local) equilibrium captured by hydrodynamics. This leads naturally to the question of anomalous transport  in situations far from equilibrium. Especially with view on heavy ion physics this is not only of high theoretical interest but promises also to be of central importance for interpretation of experimental data. 

Theoretical studies of far from equilibrium behavior of strongly correlated quantum systems can be modeled efficiently with holographic methods \cite{Chesler:2013lia}. Holography is therefore also an ideal tool to theoretically investigate anomaly induced transport far from equilibrium. Up to now the focus has been on the chiral magnetic effect\cite{Lin:2013sga,Ammon:2016fru,Landsteiner:2017lwm, Fernandez-Pendas:2019rkh}. Field theoretic studies of out-of-equilibrium chiral magnetic effect based on a variety of methods have been presented in \cite{Buividovich:2016ulp, Mace:2016shq, Kharzeev:2020kgc}. Out-of-equilibrium anomalous transport in a gradient expansion in kinetic theory has bee studied in \cite{Hidaka:2018ekt}. In contrast in the present study we focus on the chiral vortical effect. One particular interesting aspect is that the chiral vortical effect receives a contribution from a mixed gauge-gravitational Chern-Simons term that can be interpreted as the holographic dual of the gravitational contribution to the axial anomaly \cite{Landsteiner:2011cp}. Further studies of the connection between the chiral vortical effect and the gravitational contribution to the chiral anomaly have been presented in \cite{Jensen:2012kj, Stone:2018zel, Golkar:2015oxw}.

We will follow in our setup the previous study \cite{Fernandez-Pendas:2019rkh} based on AdS Vaidya type metrics. A holographic model based on Vaidya type metrics can be used to study the out of equilibrium CVE as well. The metric is sourced by infalling null dust (with rotation) and the linear response due to the anomalies (the Chern-Simons) terms can be calculated  from well-known holographic methods. While this is not of prime interest for possible applications to heavy ion physics we also introduce a momentum relaxation parameter via a linear scalar field background \cite{Andrade:2013gsa}. 
In (local) equilibrium the anomaly induced currents are dissipationless and therefore are not affected by the translation symmetry breaking \cite{Stephanov:2015roa,Copetti:2017ywz}. Translation symmetry breaking does however affect the anomaly induced transport out of equilibrium. In particular translation symmetry implies conservation of the energy-momentum tensor. For the momentum this simply means that it does not change if no additional momentum is injected. In general the momentum density and energy current have dissipative and anomaly induced non-dissipative contributions. If translation symmetry is broken the dissipative part will eventually vanish leaving only the anomaly induced part. 
As in \cite{Fernandez-Pendas:2019rkh} this allows to study non-trivial response in the energy-momentum currents. 

The paper is organized as follows. In section 2 we introduce the model and discuss its properties. In section 3 we briefly recall the (near) equilibrium hydrodynamics of the CVE. In section 4 we present the results for the out-of-equilibrium response as computed in our holographic model. We emphasize the role of the gravitational contribution to the anomaly and the subtle interplay of the pure gauge and the gravitational contributions. We summarize our findings and present conclusion in section 5. 
 

\section{Holographic Model}

The holographic model is constructed following a \textit{bottom-up} approach. The dual field theory is assumed to have two abelian symmetries which we denote by $U(1)_A \times U(1)_V$. In our  gravity side this is achieved via the inclusion of two abelian gauge fields $A$ and $V$, with field strengths $F=dV$ and $F_5=dA$ respectively. The model also contains Einstein gravity with negative cosmological constant.

The action will have the usual Maxwell kinetic terms plus some Chern-Simons terms accounting for the anomalies. These terms are topological and gauge invariant up to a total derivative. Therefore, under a gauge transformation they give a non-trivial boundary contribution which reproduces the chiral anomaly on the boundary. We will work with a consistent form of the anomaly that preserves the classical Ward identities for the energy-momentum tensor and the vector current. In other words, we are shifting all the vector and gravitational anomalies into the axial part\footnote{Also referred to as mixed gauge-gravitational anomaly. In a quantum field theory one can always add convenient Bardeen counterterms to get a classically conserved energy-momentum tensor \cite{AlvarezGaume:1983ig}.}. We also include three massless scalar fields to allow for translation symmetry breaking \cite{Andrade:2013gsa}.
The action of our model is 

\begin{equation}
\begin{split}
    S=&\dfrac{1}{2\kappa^2} \int_{\mathcal{M}} d^5x\sqrt{-g}\left[R + \dfrac{12}{L^2}-\dfrac{1}{2}\partial_\mu X^I\partial^\mu X^I -\dfrac{1}{4}F^2-\dfrac{1}{4} F_5^2 \right. \\  & \left.+ \dfrac{\alpha}{3}\epsilon^{\mu \nu \rho \sigma \tau} A_\mu \left( 3F_{\nu \rho}F_{\sigma \tau}+F^5_{\nu \rho}F^5_{\sigma \tau} \right)+ \lambda \epsilon^{\mu \nu \rho \sigma \tau} A_{\mu}R^\alpha \s _{\beta \nu \rho} R^\beta \s _{\alpha \sigma \tau} \vphantom{\left[\dfrac{12}{L^2} \right.}\right] \\ & + \dfrac{1}{\kappa^2} \int_{\partial \mathcal{M}} d^4x\sqrt{-\gamma} K + S_{nf}  ,
\end{split}
\end{equation}

\noindent
where $I=1,2,3$, $\kappa^2$ is the Newton constant, $L$ the AdS radius, and $\alpha$, $\lambda$ are the Chern-Simons couplings. The Levi-Civita tensor is defined as $\epsilon^{\mu \nu \rho \sigma \tau}=\epsilon(\mu \nu \rho \sigma \tau)/\sqrt{-g}$. The boundary term corresponds to the Gibbons-Hawking action, which is required to have a well-defined variational problem\footnote{One could also introduce a Chern-Simons boundary term depending on the extrinsic curvature. In asymptotically AdS spaces this terms does not contribute to the on-shell action. Therefore we do not include it here \cite{Landsteiner:2011iq}}. As usual $\gamma$ is the induced metric on $\partial \mathcal{M}$ and $K$ the extrinsic curvature. Finally, $S_{nf}$ represents the contribution of a null fluid sourcing the  Vaidya-like solutions.\\

In order to allow for non-trivial response in the energy-current we should break translational invariance. Also in a view of possible applications to condensed matter physics translation symmetry breaking is a more realistic scenario. The free scalar fields will be given a profile of the form $X^I=k x^i$ for some constant $k$, where $x^i\in \{x,y,z\}$ are the three spatial coordinates. Because the scalar fields only enter through derivatives the background will still be translation invariant. In particular the metric and gauge field backgrounds will be $x^i$ independent. Translation breaking will be manifest however in higher point correlation function or equivalently by probing the background with perturbations such as the magnetic field or by rotation. 
In this construction the graviton will acquire an effective mass\cite{Blake:2013owa}. The parameter $k$ can be thought of as encoding the presence of a uniformly distributed density of impurities in the dual field theory.

The corresponding equations of motions are 

\begin{equation}
    2\kappa^2 Y^I_{(nf)}=\dfrac{1}{\sqrt{-g}}\partial_{\mu}(\sqrt{-g}\partial^{\mu}X^I)
\end{equation}
\begin{equation}
\label{eq:2.3}
    2\kappa^2J^{\mu}_{(nf)}=\nabla_\nu F^{\nu \mu} +2 \alpha \epsilon^{\mu \nu \rho \sigma \tau} F_{\nu \rho }F^5_{\sigma \tau}
\end{equation}
\begin{equation}
\label{eq:2.4}
\begin{split}
    2\kappa^2J^{\mu}_{5,(nf)}=&\nabla_\nu F_5^{\nu \mu} + \alpha \epsilon^{\mu \nu \rho \sigma \tau}( F_{\nu \rho }F_{\sigma \tau}+F^5_{\nu \rho }F^5_{\sigma \tau})\\&+ \lambda \epsilon^{\mu \nu \rho \sigma \tau} R^\alpha \s _{\beta \nu \rho} R^\beta \s _{\alpha \sigma \tau}
\end{split}
\end{equation}
\begin{equation}
    \begin{split}
        \kappa^2 T^{(nf)}_{\mu \nu}=& G_{\mu \nu} -\dfrac{6}{L^2}g_{\mu \nu} - \dfrac{1}{2}\partial_\mu X^I\partial_\nu X^I +\dfrac{1}{4} \partial_\rho X^I \partial^\rho X^I g_{\mu \nu}\\&-\dfrac{1}{2}F_{\mu \rho} F_{\nu}\s^{\rho}  +\dfrac{1}{8}F^2g_{\mu \nu} -\dfrac{1}{2}F^5_{\mu \rho} F^5_{\nu}\s^{\rho} +\dfrac{1}{8}F_5^2g_{\mu \nu}\\&-2\lambda \epsilon_{\lambda\rho \sigma \tau (\mu |} \nabla_{\beta} \left[ F_5^{\rho \lambda} R^\beta \s _{|\nu)}\s ^{\sigma \tau} \right].
    \end{split}
\end{equation}

We are ultimately concerned with computing the currents and the energy momentum tensor in the quantum field theory. Those can be obtained through the standard holographic prescription: varying the on-shell action with respect to the boundary value of the dual field \cite{Zaanen:2015oix,Ammon:2015wua}

\begin{equation}
    Y^I= \left. \sqrt{-\gamma} n_{\mu} \partial^{\mu} X^I\right|_{\partial\mathcal{M}}
\end{equation}

\begin{equation}
\label{eq:2.8}
    J^\alpha=\left. \sqrt{-\gamma}n_{\mu} \left[ F^{\alpha\mu}+4\alpha \epsilon^{\mu \alpha\beta\gamma\delta}A_\beta F_{\gamma\delta} \right]\right|_{\partial \mathcal{M}}
\end{equation}

\begin{equation}
\label{eq:2.9}
    J^\alpha_5= \left. \sqrt{-\gamma} n_{\mu} \left[ F^{\alpha\mu}_5 +\dfrac{4\alpha}{3}\epsilon^{\mu \alpha\beta\gamma\delta}A_\beta F_{\gamma\delta}^5 \right]\right|_{\partial \mathcal{M}}
\end{equation}

\begin{equation}
\label{eq:2.10}
    T^{\alpha\beta}=\left. 2\sqrt{-\gamma}\left[ -K^{\alpha\beta}+\gamma^{\alpha\beta}K-4\lambda n_{\mu} \epsilon^{\mu (\alpha|\gamma\delta\epsilon} \left( \dfrac{1}{2} F_{\gamma\delta}\tilde{R}^{|\beta)}\s _{\epsilon} +D_{\eta} A_\gamma\tilde{R}^{\eta|\beta)}\s _{\delta\epsilon} \right) \right]\right|_{\partial \mathcal{M}},
\end{equation}

\noindent
with $n_{\mu}$ a normalized orthonormal vector to the surface that defines the boundary \footnote{We note that there are additional terms in the holographic stress tensor depending on the extrinsic curvature stemming from the gravitational Chern-Simons term. These terms have been identified in \cite{Copetti:2017ywz,Copetti:2017cin}. They contribute to the chiral magnetic response in the energy-current. One can check that these terms do not contribute to the chiral vortical response in our setup.}.
$\tilde R$ is the intrinsic curvature the boundary and $D$ the intrinsic covariant derivative. Note that the action has not been renormalized and appropriate counterterms have to be added to get finite expression. As we will see the chiral vortical responses are finite by themselves.   

\subsection{Background Solution}

We want to explore the out of equilibrium behavior of the CVE in the linear response regime and in the presence of momentum relaxation. Our background must then represent a thermal time-evolving homogeneous and isotropic charged state. The rotation will be included later as a perturbation on top of this background. The simplest setup satisfying these requirements is a black brane with time dependent blackening factor. In Eddington-Finkelstein coordinates we have
\begin{equation}
    ds_0^2=-f(v,r)dv^2+2drdv+r^2\delta_{ij}dx^idx^j.
\end{equation}
The boundary is located at $r\to \infty$ and the apparent horizon at some $r=r_H$ which is the largest real root of $f(v,r)$. For numerical convenience we work with the variable $u=1/r$.

Our background solution should also have a non-vanishing chemical potential. 
We therefore take the  field strengths to be of the form $F_{uv}=-uq$ and $F^5_{uv}=-uq_5$. We will work in the radial gauge $A_u=V_u=0$. The equations of motion \ref{eq:2.3} and \ref{eq:2.4} reveal that charges can only vary with time via an external source
\begin{equation}
    2\kappa^2 J^u_{(nf)}=-\Dot{q} u^5~~,~~2\kappa^2 J^u_{5,(nf)}=-\Dot{q}_{5} u^5
\end{equation}
where we represent $v$ derivatives with a dot. A Vaidya-like solution for the blackening factor is found analytically:
\begin{equation}
\label{eq:2.13}
    f(v,u)=\dfrac{1}{L^2u^2}-\dfrac{1}{4}k^2-2m(v)u^2+\dfrac{1}{12}[q(v)^2+q_5(v)^2]u^4,
\end{equation}
where we can set $L$ to $1$ without altering the physics. The Einstein equations are sourced by the bulk energy-momentum tensor of infalling null dust
\begin{equation}
    k^2T_{vv}^{(nf)}=3\Dot{m}u^3-\dfrac{q\Dot{q}}{4}u^5-\dfrac{q_5\Dot{q}_5}{4}u^5.
\end{equation}
The chemical potential of the field theory is dual to the temporal component of the gauge field at the boundary minus its value at the horizon, whereas the temperature is that of the black brane. Hence,

\begin{equation}
\label{eq:II.14}
    \mu = \dfrac{q}{2}u_H^2~~,~~\mu_{5} = \dfrac{q_{5}}{2}u_H^2
\end{equation}
and
\begin{equation}
\label{eq:2.16}
\begin{split}
    T=& \left. \dfrac{1}{2\pi}\left( -\dfrac{u^2}{2}\partial_u f(v,u) \right) \right|_{u=u_H}\\=&\dfrac{1}{4\pi}\left( \dfrac{k^2}{2}u_H+8mu_H^3 -\dfrac{1}{2}(q^2+q_5^2)u_H^5 \right),
\end{split}
\end{equation}

\noindent
where we have used that by definition $f(v,u=u_H)=0$. We stress that these expression can be interpreted as chemical potentials and temperatures only in equilibrium. It is nevertheless useful to define them in this way since the form of the background metric and gauge fields is suggestive of instantaneous thermalization. We can then compare the hydrodynamic response based on these definitions take it as a benchmark for the actual out-of-equilibrium response.

\subsection{Linear Perturbation Regime}

The chiral vortical effect is a current generated in a rotating system as a consequence of the chiral anomaly. Modeling it in holography requires to encode rotating boundary conditions in the gravity dual. This has been successfully achieved via the inclusion of a mixed time-spatial component in the metric, corresponding to slowly rotating objects which induce by themselves rotation through the frame dragging effect. In particular we introduce a gravito-magnetic field via
\begin{equation}
    dv \to dv+\epsilon \Vec{A}_g(v,x,y,z)d\Vec{x}
\end{equation}
in \ref{eq:2.13}. We choose the vector to be $\Vec{A}_g=(B_g y,0,0)$ without loss of generality. In the gravito-magnetic formalism $\Vec{A}$ is regarded as a gauge field and our choice generates a gravito-magnetic field $\Vec{B_g}=(0,0,B_g)$. One can construct a precise map to a classical rotating system through $2\Vec{\omega}=\Vec{B}_g$ with $\vec \omega$ the angular velocity. Upon the above shift, the metric turns into 
\begin{equation}
    ds^2=ds_0^2 -2\epsilon f(v,u) \,B_g y\,  dvdx - \dfrac{2\epsilon}{u^2}\,B_g y \,dudx + \cdots
\end{equation}
The dots containing higher order terms in $\epsilon$. Let us now switch on the minimal set of fluctuations required by consistency.\\





A charged rotating system will by itself present a current following the movement of the particles due to that rotation, even if no anomaly is present, and a magnetic field should be induced perpendicular to the plane of rotation. This classical behavior can be parametrised with a perturbation in the gauge fields 
\begin{align}
    \delta A_{x}&=\epsilon B_g y a_{x}(v,u)\,,\\
    \delta V_{x}&=\epsilon B_g y v_{x}(v,u)\,.
\end{align}
We have used the notion of classical physics to factor out the $y$ dependence.

We also count with the ad-hoc knowledge that a rotation on the $x-y$ plane combined with the pure chiral anomaly should generate a time and energy dependent current in the $z$ direction, which can be accounted for with another perturbation of the gauge field and of the metric.
\begin{align}
    \delta A_{z}&=\epsilon a_z(v,u)\,,\\
    \delta V_{z}&=\epsilon v_z(v,u)\,,\\
    \delta g_{vz}&=\epsilon \dfrac{h(v,u)}{u^2}\,.
\end{align}
Finally, one should notice that this modification of spacetime prevents us for giving the simple profile to the scalar field $X^3$ that we were considering before. The condition is too stringent and one should allow the field to vary with $v$ and $u$ at $\mathcal{O}(\epsilon)$. 
\begin{equation}
    \delta X_3=\epsilon Z(v,u).
\end{equation}
Recall that the scalar fields were breaking translation invariance and thus effectively damping momentum through a homogeneous distribution of impurities. The inclusion of this new term necessarily has a different effect, as it does not modify the explicit breaking. The perturbation may be seen then as the effective dynamics for the impurities in the presence of rotation. This is far from a rigorous statement but it is nice to have an intuitive picture of the situation. 

Combining this information with the equations of motion for the $x$ component of the gauge fields we find a simple solution for $\delta A_{x}$ and $\delta V_{x}$, namely 
\begin{align}
\label{eq:2.26}
    \delta A_{x}=-\dfrac{u^2q_5}{2} \epsilon B_g y \,,\\
\label{eq:2.27}
    \delta V_{x}=-\dfrac{u^2q}{2} \epsilon B_g y\,.
\end{align}
Allowing the charges to vary with time requires once again the inclusion of external sources:
\begin{align}
    2\kappa^2J_{5,(nf)}^{x}&=\dfrac{\Dot{q}_5 u^5}{2}\epsilon B_g y\,,\\
    2\kappa^2J_{(nf)}^{x}&=\dfrac{\Dot{q} u^5}{2}\epsilon B_g y\,.
\end{align}
The $vx$ component for the gravitational field equations are sourced by the external energy-momentum tensor
\begin{equation}
\label{eq:2.31}
    \kappa^2 T_{vx}^{(nf)}=\epsilon B_g y \left( 4\Dot{m}u^3-\dfrac{q_5\Dot{q}_5}{4}u^5-\dfrac{q\Dot{q}}{4}u^5 \right).
\end{equation}
We interpret these as the effects on frame dragging on the infalling null dust. This  is not surprising, in order to source a Vaidya type metric with rotation the infalling null dust also has to be co-rotating.

After these preliminaries the dynamics of the system can be obtained. Substitution of previous results gives the following set of differential equations 
\begin{align}
    d'Z-\dfrac{3}{2u}dZ+\dfrac{3uf}{4}Z'+\dfrac{3k}{2u}h-\dfrac{k}{2}h'&=0 \,,
\label{eq:2.32}\\
    d'v_z-\dfrac{1}{2u}dv_z+\dfrac{uf}{4}v_z'+\dfrac{1}{2}uqh'-4\alpha B_g u^4 qq_5&=0\,,
\label{eq:2.33}
\end{align}
\begin{equation}
\begin{split}
    &d'a_z-\dfrac{1}{2u}da_z+\dfrac{uf}{4}a_z'+\dfrac{1}{2}uq_5h'-2\alpha B_g u^4 (q^2+q_5^2)-  B_g \lambda u \left[(u^2f)'^2\right]' =0,
\end{split}
\end{equation}
\begin{equation}
\label{eq:2.34}
\begin{split}
    &d'h+\dfrac{5uf+u^2f'}{2}h'-q_5u^3da_z-qu^3dv_z-kdZ+k^2h\\-&B_g\lambda u^5 q_5\left[ -12\Dot{f}-12u\Dot{f}'- 2u^2\Dot{f}'' + uf(12f+24uf'+10u^2f''+u^3f''') \right] \\ + & 2 B_g \lambda u^5 \Dot{q}_5 \left[ 6f + 6uf' + u^2f'' \right]  =0,
\end{split}
\end{equation}
supplemented with the time independent constraint 
\begin{equation}
\label{eq:consraint}
    -\left(\dfrac{h'}{u^3}\right)'+q_5a_z'+qv_z'-\dfrac{k}{u^3}Z'-2B_g q_5\lambda u [12f+24uf'+10u^2f''+u^3f''']=0.
\end{equation}
Derivatives with respect to $u$ are denoted with a prime. Time derivatives have been replaced by a directional derivative on the direction of ingoing null geodesics that simplifies the form of the equations of motion. In particular 
\begin{equation}
    d=\partial_v-\dfrac{u^2f}{2}\partial_u,
\end{equation}
and accordingly
\begin{equation}
    d'=\partial_{vu}^2-\dfrac{u^2f}{2}\partial_{uu}^2-uf\partial_u-\dfrac{1}{2}u^2f'\partial_u.
\end{equation}
The whole system of equations and constraints needs to be consistent. One can check that 
\begin{equation}
\begin{split}
    -\Dot{q}_5a_z'-\Dot{q}v_z'=&d(\textrm{eqII.33})-\partial_u\left(\dfrac{(\textrm{eqII.32})}{u^3}\right)\\&-\partial_u\left[\left(\dfrac{u^2f}{2}\right)(\textrm{eqII.33})\right]-\dfrac{2k}{u^3}(\textrm{eqII.29}),
\end{split}
\end{equation}
which should vanish exactly on shell. In general this can only happen if both charges are constant in time.
As in \cite{Fernandez-Pendas:2019rkh} we will from now on restrict to solutions with $\dot{q}=\dot{q}_5 =0$.

\newpage

\section{Near Equilibrium Chiral Vortical Effect}

The equilibrium configuration will provide us with the initial profiles of the fields. We shall assume that the system is originally at thermal equilibrium and perform a quench in the mass. The evolution of the system can be compared to that of hydrodynamics, so that deviations from it signal out out of equilibrium behavior.

\subsection{Hydrodynamics}
The constitutive relations for the currents are 
\begin{align}
    J^{\mu} &= \rho u^{\mu} + \sigma \omega^\mu\,,\\
    J_5^{\mu} &= \rho_5 u^{\mu} + \sigma_5 \omega^\mu\,,\\
    T^{\mu \nu} &= (\epsilon + p)u^\mu u^\nu + p \eta^{\mu \nu} + \xi (u^\mu \omega^\nu + u^\nu \omega^\mu)\,.
\end{align}
Usually, in relativistic hydrodynamics, there is an ambiguity in the choice of frame, which gets mapped into an ambiguity in the choice of boundary conditions in holography. However, regularity of the solution in the horizon imposes one more condition: the metric perturbation $h(v,u)$ has to vanish in the horizon. The system chooses thus a preferred frame. This only happens when $k\neq 0$, so we can state it is a consequence of momentum relaxation in this model \cite{Copetti:2017ywz}. In a completely analogous way this also happens in a purely hydrodynamic model with impurities \cite{Stephanov:2015roa}. 
Physically there is a preferred frame in which the impurities are at rest. In this frame the coefficients are
\begin{align}
\label{eq:3.4}
    \sigma &= 16 \alpha \mu \mu_5\,,\\ 
\label{eq:3.5}
    \sigma_5 &= 8 \alpha (\mu^2 + \mu_5^2) + 64 \pi^2 \lambda T^2\,,\\
\label{eq:3.6}
    \xi &= \dfrac{16}{3} \alpha \mu_5 (3 \mu^2 + \mu_5^2 ) + 128 \pi^2 \lambda \mu_5 T^2\,.
\end{align}
Both axial and energy currents acquire a quadratic temperature dependence directly related to the mixed gauge-gravitational anomaly. Deviations from these 1-point functions will be taken as signature of out of equilibrium dynamics.\\  

The leading terms in the hydrodynamic expansion also deserve further attention. These describe a convective flow which is subject to momentum relaxation so long as $k \neq 0$. Consequently, its contribution eventually fades and the final equilibrium state is described in terms of \ref{eq:3.4}-\ref{eq:3.6} only. The situation dramatically changes for $k=0$, i.e. for unbroken translation symmetry. Now momentum is conserved and the convective flow arising from the dynamics of the system will not vanish, modifying the near equilibrium evolution and, in particular, the final equilibrium state. We can exploit momentum conservation to find the fluid velocity. We specialize here to the case where the rotation is around the $z$ axis:
\begin{equation}
    \left<T^{0z}\right>_{in}=\xi_{in} \omega^z = (\epsilon + p)v_z + \xi \omega^z = \left<T^{0z}\right>\, ,
\end{equation}
so that
\begin{equation}
    v_z = \dfrac{\omega^z}{(\epsilon+p)}(\xi_{in}-\xi)\, .
\end{equation}
Therefore, both the axial and vector currents become 
\begin{equation}
    J^{z}_{(5)}=\dfrac{\rho_{(5)}\omega^z}{(\epsilon+p)}(\xi_{in}-\xi) + \sigma_{(5)}\omega^z\, .
\end{equation}
Finally, we resort to the holographic dictionary to determine
\begin{equation}
    \epsilon=\left<T^{00}\right>=6 m\, ,
\end{equation}
\begin{equation}
    p=\left<T^{ii}\right>= 2m\, ,
\end{equation}
\begin{equation}
    \rho_{(5)}=\left<J^0_{(5)}\right>=q_{(5)}\, .
\end{equation}
\subsection{Boundary Behavior}


According to the holographic dictionary the vacuum expectation values of the operators in the dual field theory can be read off from the asymptotic expansion of the AdS bulk fields near the boundary ($u \to 0$). The leading terms in this expansion are interpreted as the sources for the operators. The expectation values of the operators are related to the subleading terms in the expansion. In our problem we are only interested in the expectation values of the currents, and we do not want our operators to be sourced. Therefore we set the leading boundary modes to zero.  Doing the asymptotic analysis we find:
\begin{align}
\label{eq:3.7}
    v_z &\simeq V_2 u^2 + \Dot{V_2} u^3 + \mathcal{O}(u^4)\,,\\
    a_z &\simeq A_2 u^2 + \Dot{A_2} u^3 + \mathcal{O}(u^4)\,,\\
    Z &\simeq Z_4 u^4 + \dfrac{5\Dot{Z}_4-kh_4}{5}u^5 + \mathcal{O}(u^6)\,,\\
\label{eq:3.10}
    h &\simeq h_4u^4-\dfrac{4kZ_4}{5}u^5 + \mathcal{O}(u^6)\,.
\end{align}
Besides, from the constraint one finds $\Dot{h}_4=-kZ_4\,$, which is related to the non-conservation of the energy current for $k \neq 0\,$. This information can now be put into \ref{eq:2.8}-\ref{eq:2.10} to find 
\begin{align}
\left\langle J^z \right\rangle &= 2 V_2 \epsilon\,,\\
\left\langle J_5^z \right\rangle &= 2 A_2 \epsilon\,,\\ 
\left\langle T^{vz} \right\rangle &= - 4 h_4 \epsilon.
\end{align}
The subleading modes are indeed giving us the expectation values of the operators. Our goal then reduces to find the time evolution of the asymptotic coefficients.

\subsection{Equilibrium in holography}
The anomaly induced equilibrium response in holography with momentum relaxation has been studied in \cite{Copetti:2017ywz}. The response coefficients are precisely given by (\ref{eq:3.4}-\ref{eq:3.6}) and completely determined by  chemical potentials and temperature and no explicit dependence on the momentum relaxation parameter $k$.

\newpage

\section{Out of Equilibrium CVE}
In all the cases we studied  a quench on the mass of the form 
\begin{equation}
    m=m_0 + \dfrac{m_f-m_0}{2}\left( 1+\textrm{tanh}\left( \dfrac{v}{\tau} \right) \right),
\end{equation}
with the masses conveniently chosen so that at initial and final times the horizon lies at $u_h^{in}=1$ and $u_h^{fin}=0.8$ respectively. As for the topological Chern-Simons couplings we choose them so that they reproduce the chiral anomaly for a quantum field theory with $N_f$ Dirac fermions, those are $\alpha=N_f(16 \pi^2)^{-1}$ and $\lambda=N_f(384\pi^2)^{-1}\,$. At the linear level, the system is only sensitive to the ratio $\alpha/\lambda\,$, as rescalings of both coefficients at a time can be reabsorbed into $B_g\,$. For practical purposes we can effectively set $\alpha = B_g = 1$ and $\lambda=32/768\,$.


\subsection{Dependence on translation symmetry breaking}
\label{sec:4A}

\begin{figure}[h!]
    \centering
    \includegraphics[scale=0.5]{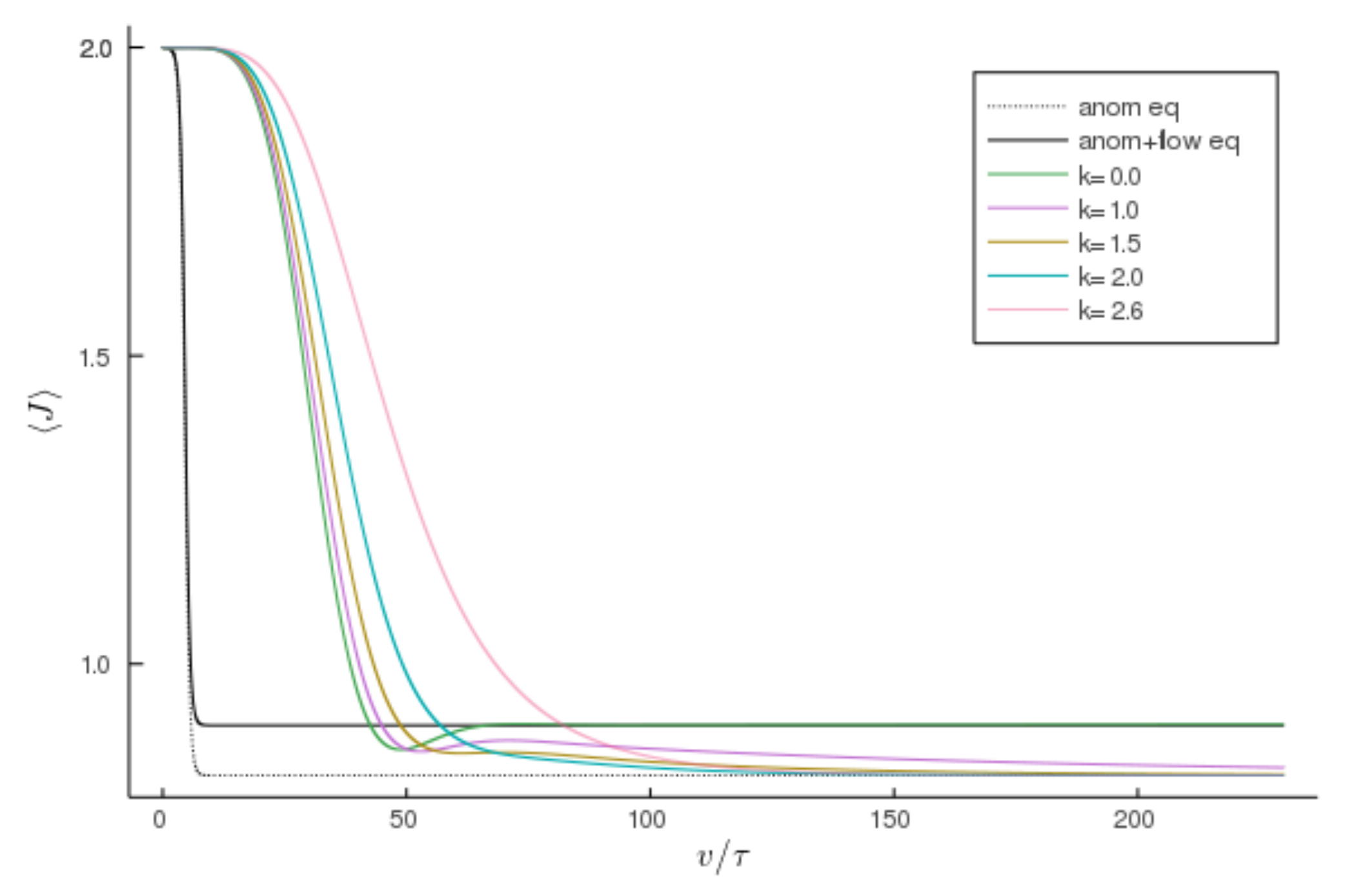}
    \caption{\small{Momentum relaxation dependence for the vector current. We fix $q=q_5=1.0$ and $\tau=0.05\,$. 
The dashed lines show the response due to the anomaly only while the continuous line is taking into account the effect of collective flow in the momentum conserving case $k=0\,$.    
    We can observe a transition between oscillating and purely damped behavior.}}
    \label{fig:1}
\end{figure}

\begin{figure}[h!]
    \centering
    \includegraphics[scale=0.45]{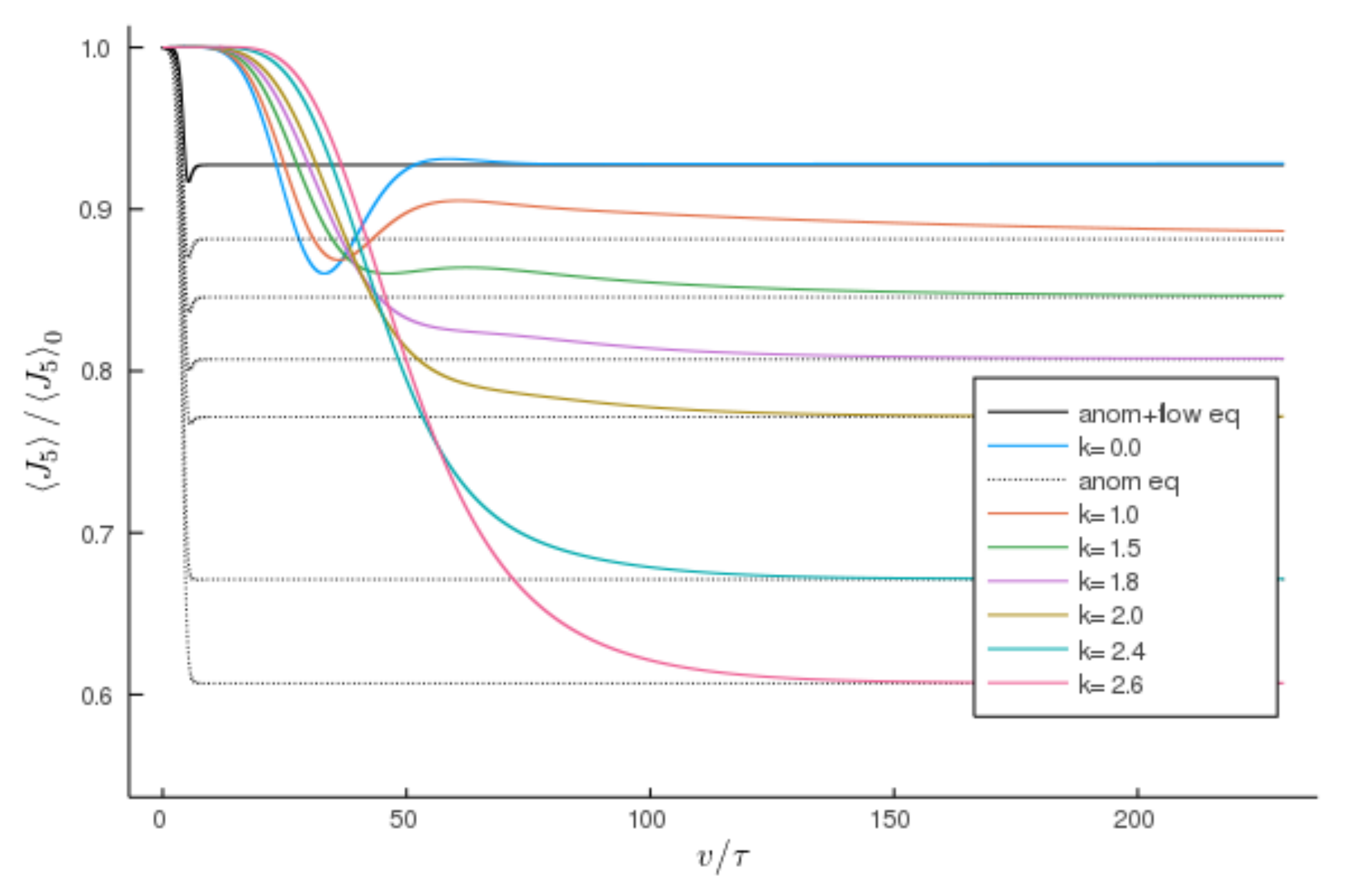}
    \caption{\footnotesize{Momentum relaxation dependence for the axial current. We plot the axial current normalized to its initial value. We fix $q=q_5=1.0$ and $\tau=0.05\,$.
Dashed lines show the anomaly response only and the continuous black line is the hydrodynamic result taking flow into account.    
     Again there is a transition between oscillating and continuously decreasing behavior. }}
    \label{fig:2}
\end{figure}

\begin{figure}[h!]
    \centering
    \includegraphics[scale=0.45]{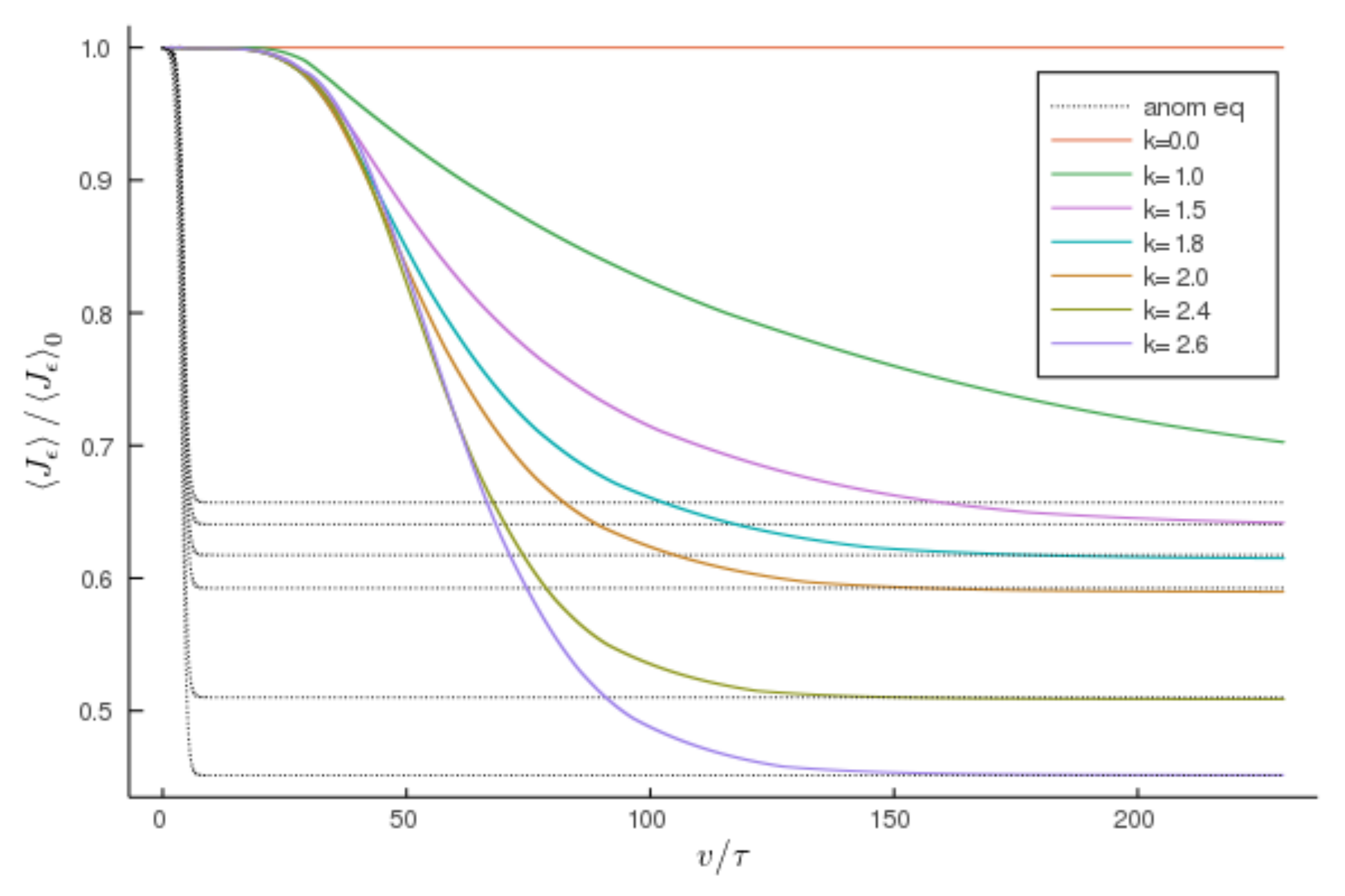}
    \caption{\footnotesize{Momentum relaxation dependence for the energy current. We fix $q=q_5=1.0$ and $\tau=0.05\,$. The current is plotted normalized to its initial value, so that one sees the percentage of energy current that remains at each time. 
    As expected the response in the momentum preserving case $k=0$ is trivial, the anomalous response is completely compensated by the dissipative flow. For $k\neq 0$ the flow eventually dissipates and only the anomalous non-dissipative response is left over for large times. Contrary to the responses in the charge currents there is no regime of oscillating behavior.}}
    \label{fig:3}
\end{figure}

In this section we keep both charges fixed to unity and study the evolution of the currents for different values of the momentum relaxation parameter $k$. The quench parameter is set to $\tau=0.05\,$. The results for vector, axial and energy current are shown in figures \ref{fig:1},\ref{fig:2} and \ref{fig:3} respectively.

When translation symmetry is not broken the convective flow does not dissipate. 
As a consequence, the response of the system in the final state is the sum of a convective term due to flow and the anomalous contribution. With translation symmetry breaking the flow will eventually stop and the currents take the values determined by the anomalies.
As functions of the translation breaking parameter $k$ the vector and axial currents (figures \ref{fig:1} and \ref{fig:2} respectively) show two qualitatively different regimes. 
For small $k$ values the current has a local minimum, raises again and then decreases to its equilibrium value.
For larger $k$ values 
the time evolution of the current is monotonically decreasing to the equilibrium value. As we will see, the study of the quasinormal modes in section \ref{sec:qnm} also suggests a transition between two regimes. Around $k\simeq 1.5$ both currents develop a plateau before finally relaxing. 
We interpret this difference as a sign that no significant flow is built up for larger $k$ values.

For the axial and energy currents the quadratic dependence on temperature has important consequences.
Since in our setup we only inject energy but charge and axial charge are conserved it means that the corresponding chemical potentials have to decrease. The temperature also depends on the momentum relaxing parameter $k$ and therefore the initial value will now also depend on it. In order to to account for this we normalize the plots for axial and energy currents to their respective initial values.

Let us concentrate now on the purely anomalous axial response denoted by the dashed lines in figure \ref{fig:2}. The axial current shows a tiny minimum in equilibrium shortly before the quench is finished. The reason is that we have increasing temperature (the infalling matter at the horizon enters the black brane increasing its temperature) and decreasing chemical potentials. The rate at which these two change is different, leading to the observed behavior. 
For the actual response including flow and its relaxation the picture is quite different. We see that the response is much slower and for the momentum preserving case the anomalous contribution is partially compensated by the flow. Again we see the transition from oscillating behavior to continually decreasing. 
In the actual response, just as in \cite{Fernandez-Pendas:2019rkh}, we have two different time scales. The first one governs the quench evolution, whereas the second one is linked to dissipation of the flow when momentum relaxation is present. 

Now let us analyze the response in the energy current \ref{fig:3}. 
First we notice that in the purely anomalous response (dashed lines) the small minimum before equilibration appearing in the axial current is absent for the energy current. This is related to the particular choice of charge values. We can check that indeed the minimum appears for a different choice of charges (see figure \ref{fig:-8}).
With no momentum relaxation the response is trivial as expected. Once translation symmetry is lost, this current relaxes faster for increasing $k$ up until $k\simeq 2.4\,$. After that it bounces back and relaxes later. As we will see this is also in agreement with the quasinormal mode analysis.

\newpage
\subsection{Quench Dependence}

\begin{figure}[h!]
    \centering
    \includegraphics[scale=0.5]{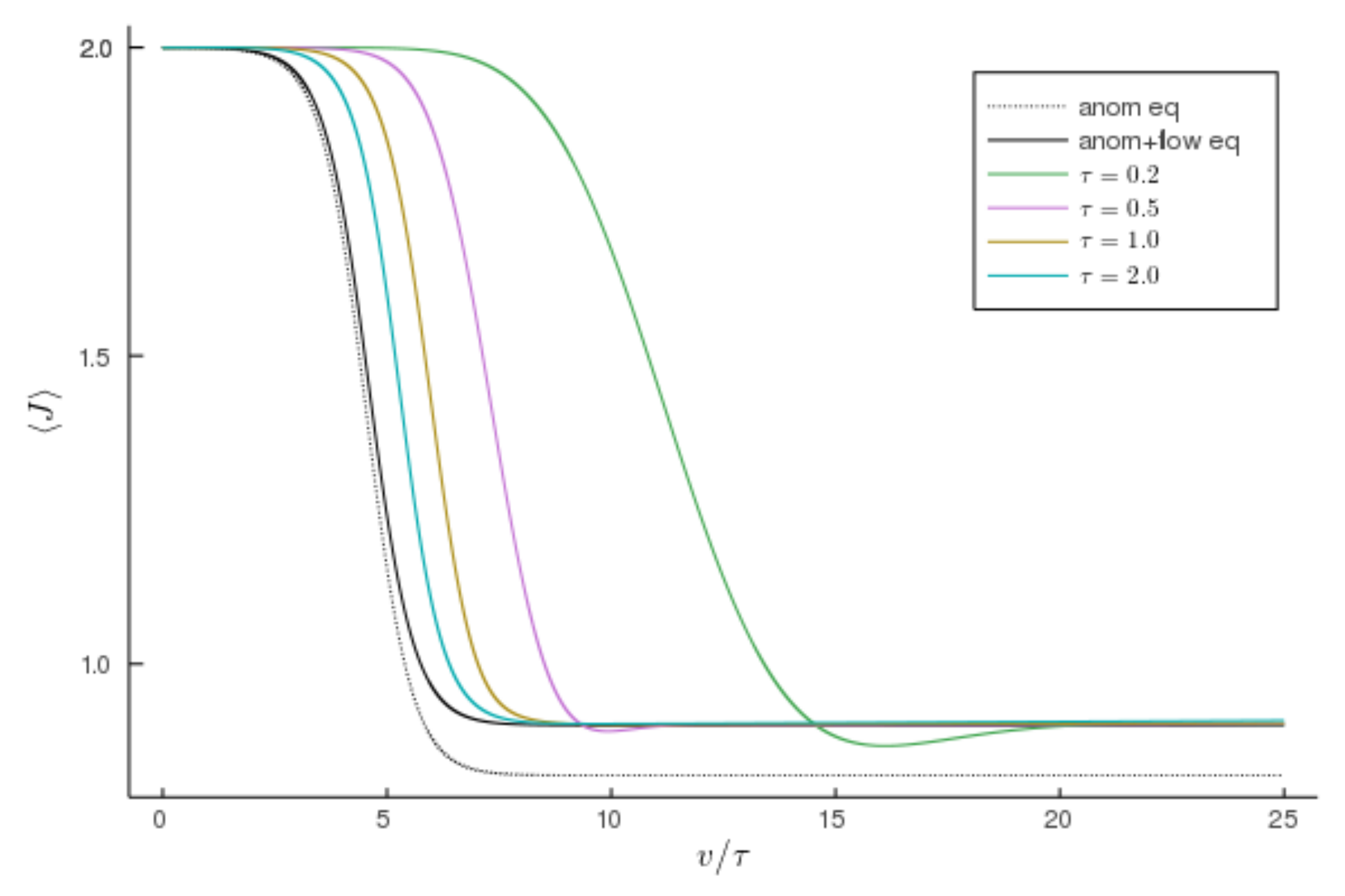}
    \caption{\small{Quench dependence for the vector current with conserved momentum. We fix $q=q_5=1.0$ and $k=0.0\,$. We observe that the slower the quench, the closer the current is to equilibrium. Momentum conservation induces a flow in the system which lifts the final stationary state.}}
    \label{fig:-1}
\end{figure}

\begin{figure}[h!]
    \centering
    \includegraphics[scale=0.51]{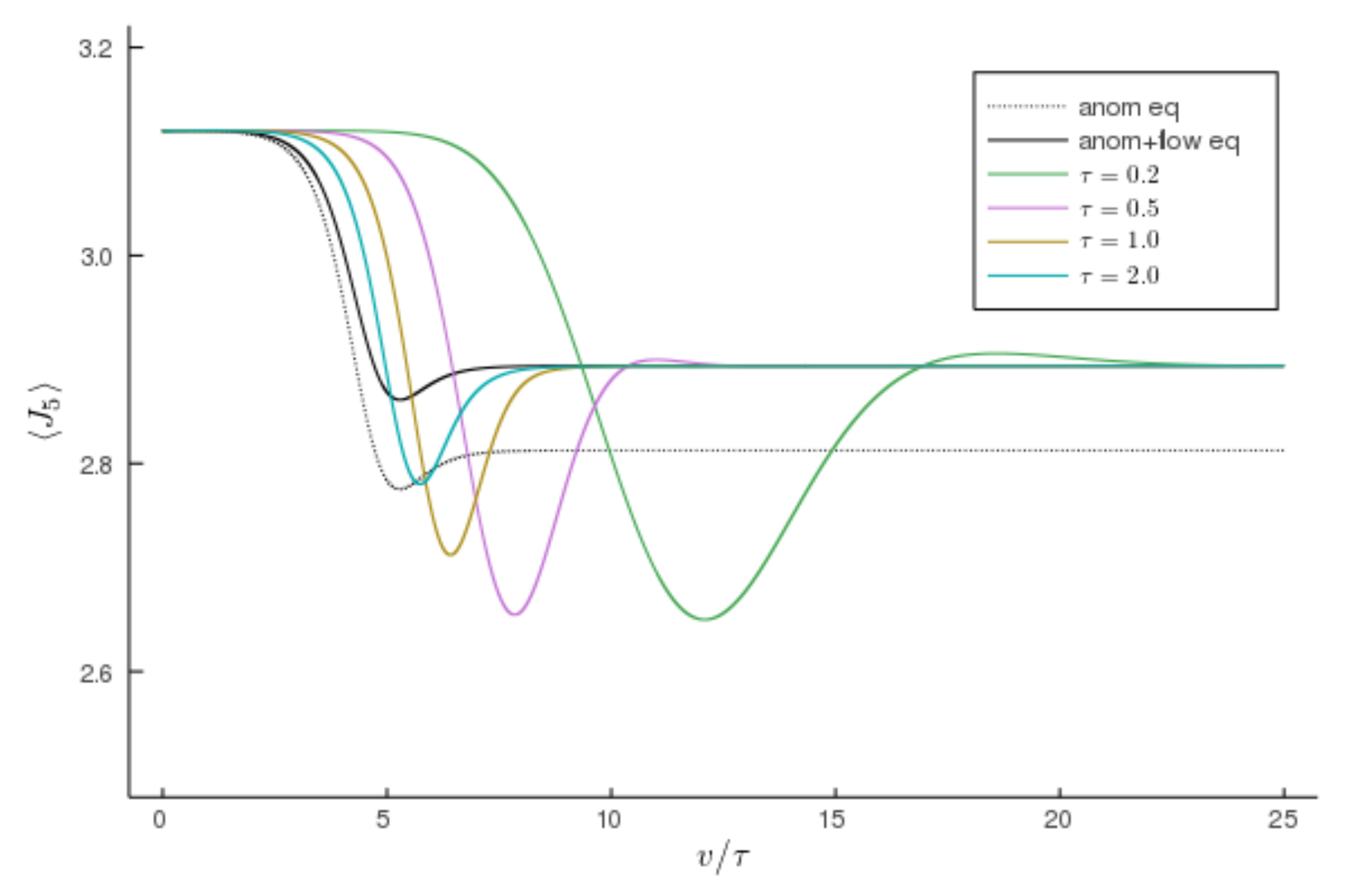}
    \caption{\small{Quench dependence for the axial current with conserved momentum. We fix $q=q_5=1.0$ and $k=0.0\,$. The interplay between the axial and gravitational anomalies leads to an absolute minimum in the current shortly before the quench has finished. This effects gets amplified out of equilibrium. Momentum conservation induces a flow in the system which lifts the final stationary state.}}
    \label{fig:-2}
\end{figure}

\begin{figure}[h!]
    \centering
    \includegraphics[scale=0.45]{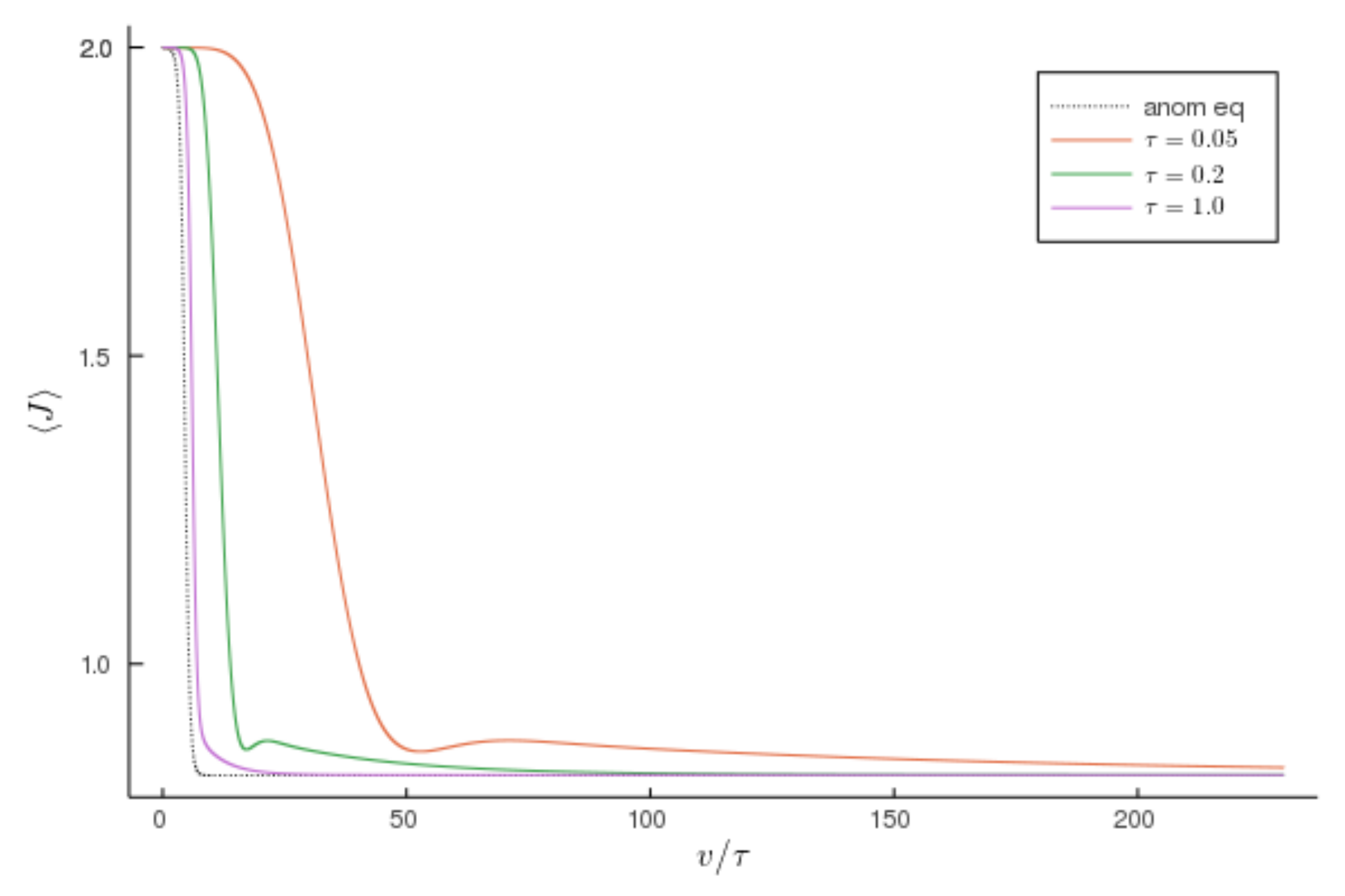}
    \caption{Quench dependence for the vector current. We fix $q=q_5=1.0$ and $k=1.0\,$. Now that momentum is no longer conserved, the current takes more time to relax. In fast quenches, say $\tau=\{0.05\,,0.2\}\,$, the current almost reaches the lifted equilibrium state of figure \ref{fig:-1}; after that momentum starts to relax. For the slow quench, the characteristic time for momentum relaxation is smaller than the timescale of the quench and so it dominates the evolution.}
    \label{fig:-3}
\end{figure}

\begin{figure}[h!]
    \centering
    \includegraphics[scale=0.45]{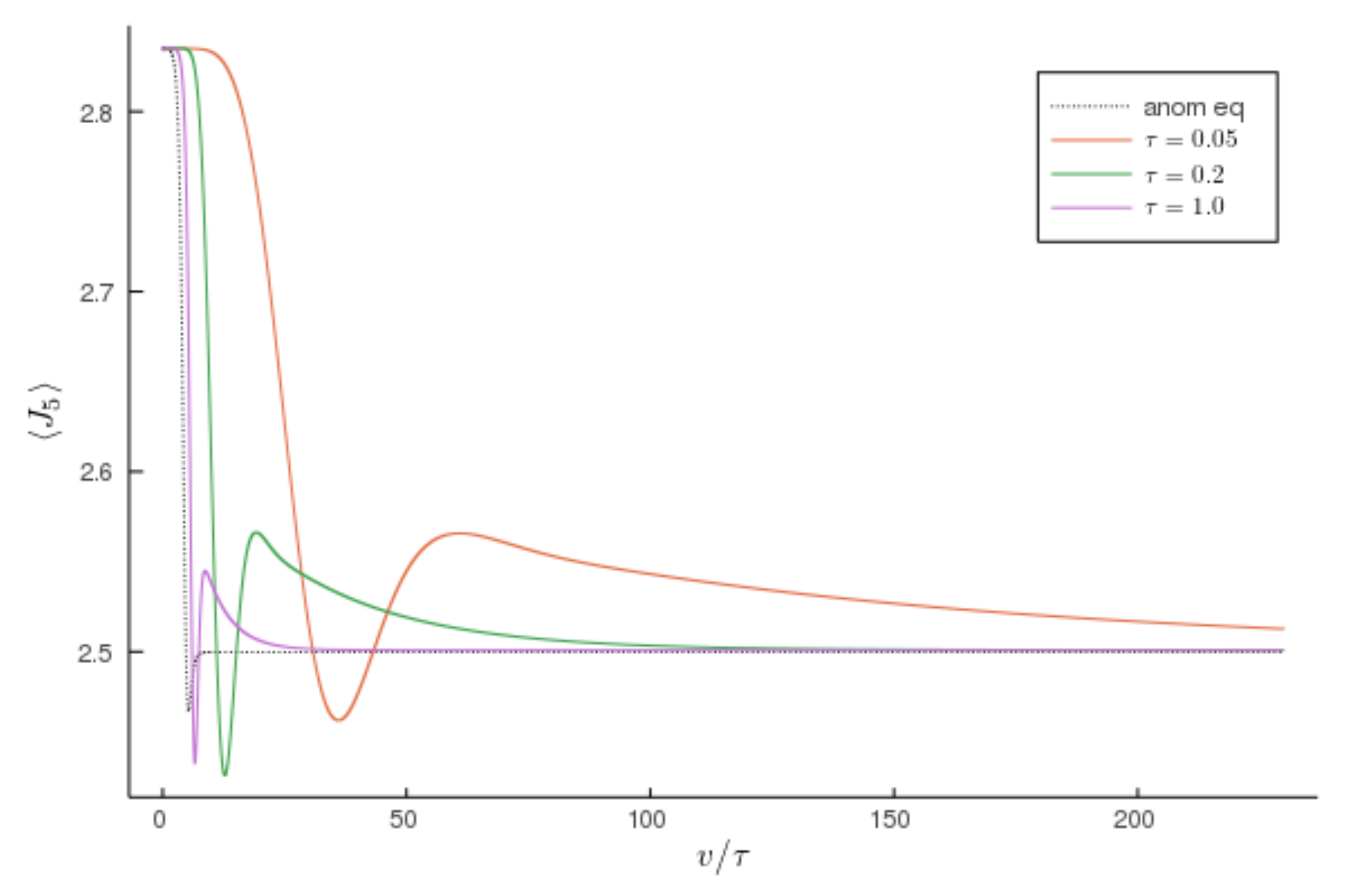}
    \caption{Quench dependence for the axial current. We fix $q=q_5=1.0$ and $k=1.0\,$. Again the equilibrium peaks get amplified. Contrary to the vector current, even the slow quench seems to get close to the stationary state with convective flow.}
    \label{fig:-4}
\end{figure}

\begin{figure}[h!]
    \centering
    \includegraphics[scale=0.5]{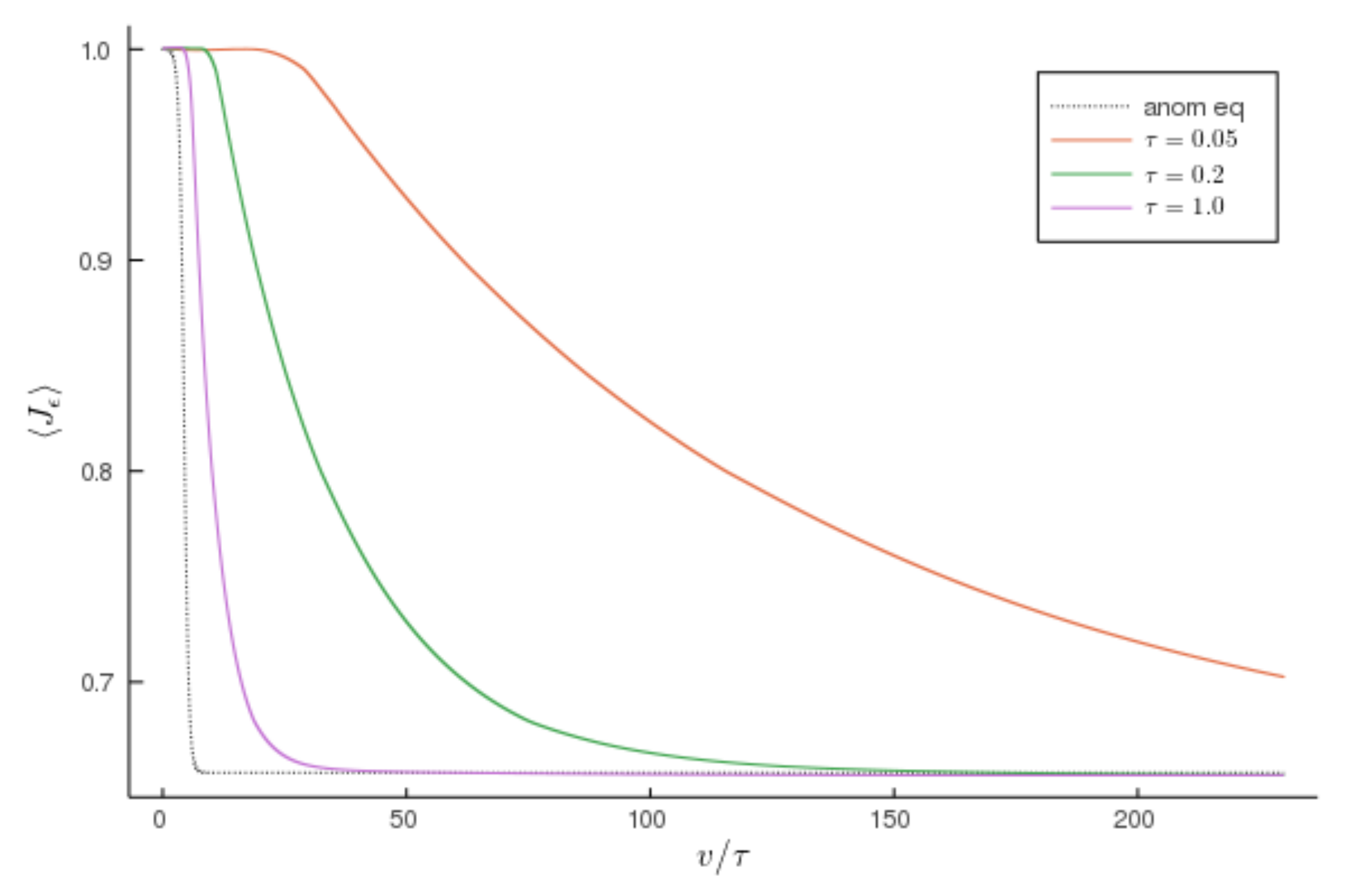}
    \caption{\small{Quench dependence for the energy current. We fix $q=q_5=1.0$ and $k=1.0\,$. The faster the quench the more the current deviates from equilibrium.}}
    \label{fig:-5}
\end{figure}

In this section we keep both charges fixed to unity and study how different quenches affect the currents. What one expects to find is that for slower quenches, i.e. larger values of $\tau$, the system never goes far from thermodynamic equilibrium at each time step and the result will approach the hydrodynamic description. For faster quenches it should deviate more and more and it is not clear a priori how it behaves in the limit where the quench is instantaneous: $\tau=0$. This will be clarified when we study the delay in the build up of the anomalous current. Our results are shown in figures \ref{fig:-1} and \ref{fig:-2} for the momentum conserving case and in figures \ref{fig:-3}, \ref{fig:-4} and \ref{fig:-5} with momentum relaxation.

We first discuss the momentum conserving case. The response pattern for vector and axial currents turn out to be quite different. The vector current never crosses below the pure anomalous result that would be obtained by ignoring the effect of the convective flow. In contrast the axial current goes considerably below this value. Note that is is also different from  what was found for the CME in \cite{Fernandez-Pendas:2019rkh}. The obvious difference is that the axial current receives the new contribution due to the the gravitational anomaly.


We have also studied the corresponding behavior with momentum relaxation. In fast quenches the currents (figures \ref{fig:-3} and \ref{fig:-4}) almost reach the equilibrium value with flow before starting to relax towards the purely anomalous equilibrium.
This is most prominent in the vector current whereas for the axial current the behavior is oscillatory even for slow quenches. Again this has to be attributed to the presence of the gravitational anomaly.

Finally the energy current is shown in figure \ref{fig:-5}. The current relaxes down to the pure anomaly induced value in monotonous fashion. 

\newpage
\subsection{Charge dependence}

\begin{figure}[h!]
    \centering
    \includegraphics[scale=0.55]{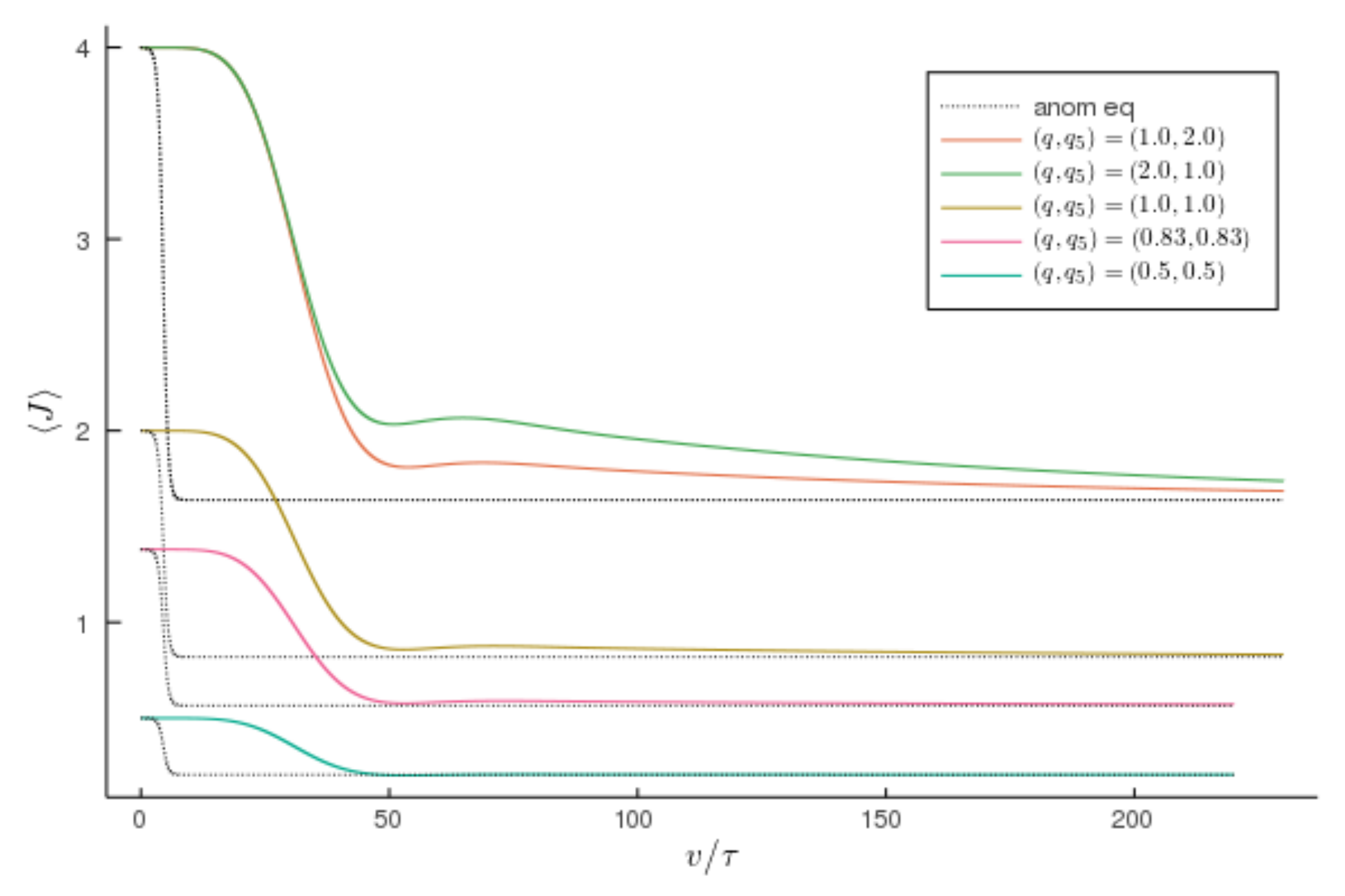}
    \caption{Vector current for various charges. We fix $\tau=0.05$ and $k=1.0$. We observe a transition between two regimes, distinguishable by the slight bounce on the current, or equivalently by the existence of a local minimum during the evolution. }
    \label{fig:-6}
\end{figure}

\begin{figure}[h!]
    \centering
    \includegraphics[scale=0.5]{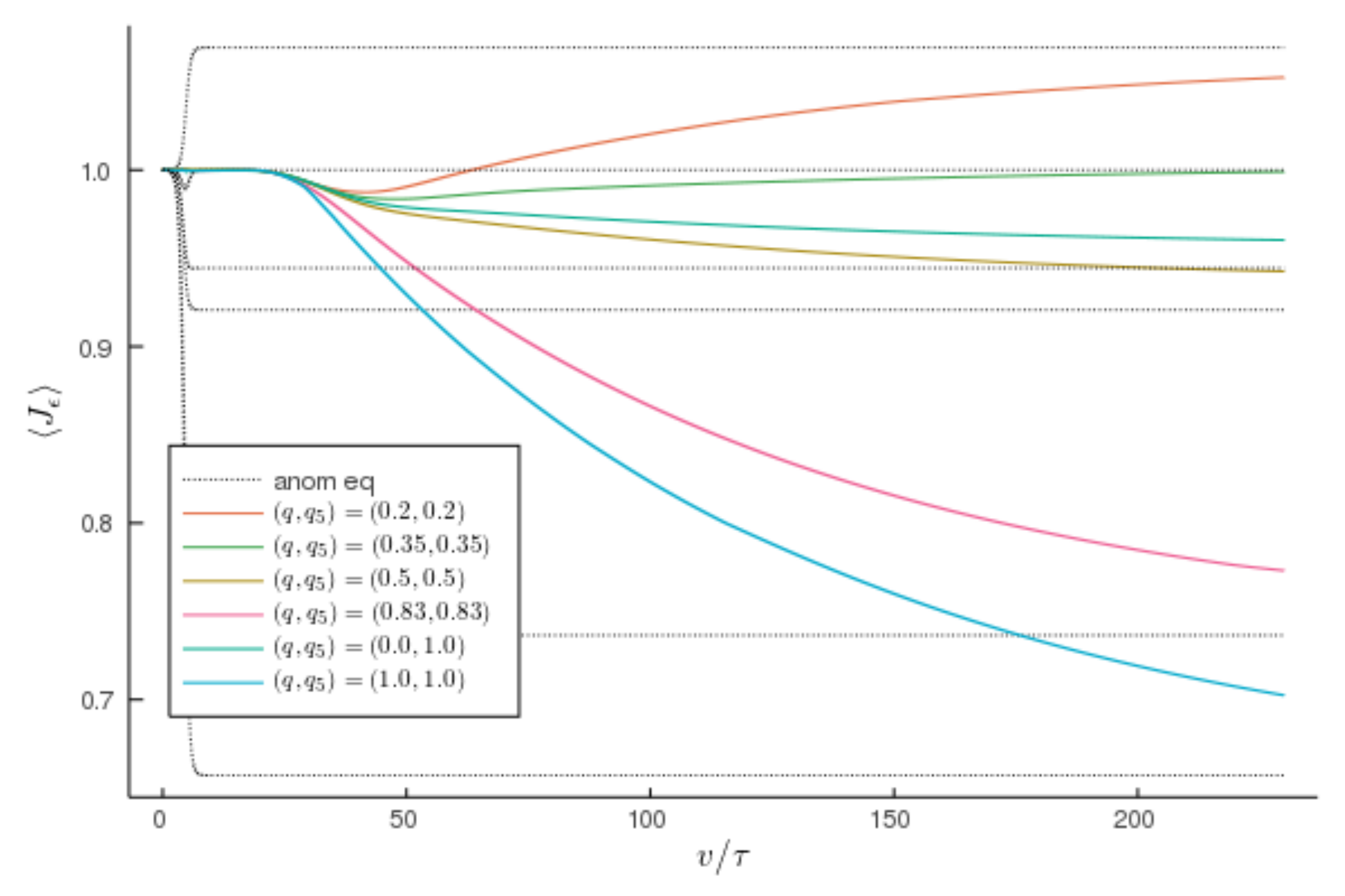}
    \caption{\small{Energy current for various charges. We fix $\tau=0.05$ and $k=1.0$. We plot the energy current normalized to its initial value. The interplay between the axial and gravitational anomalies allows for configurations where the final value of the energy current is above the initial value.}}
    \label{fig:-8}
\end{figure}

Once we have decided to perform a quench in the mass, the parameter space of the model reduces to $(\tau,k,q,q_5)$. More generically we should also include the Chern-Simons couplings in the parameter space, but we have decided to sit on the point where those reproduce the chiral anomaly as we explained at the beginning of this section. Having studied the effect of $k$ and $\tau$, we will now explore the behavior for different charges. The results are shown in figures \ref{fig:-6}, \ref{fig:-7} and \ref{fig:-8}. We begin by discussing the limiting configurations. First of all, setting $q_5=0$ one can check in the equations of motion \ref{eq:2.31}-\ref{eq:2.34} that all the perturbation except $a_z$ decouple and can be consistently set to zero. Only the axial current is present. 
On the other hand for $q=0$ only $v_z$ decouples. 

\begin{figure}[h!]
    \centering
    \includegraphics[scale=0.48]{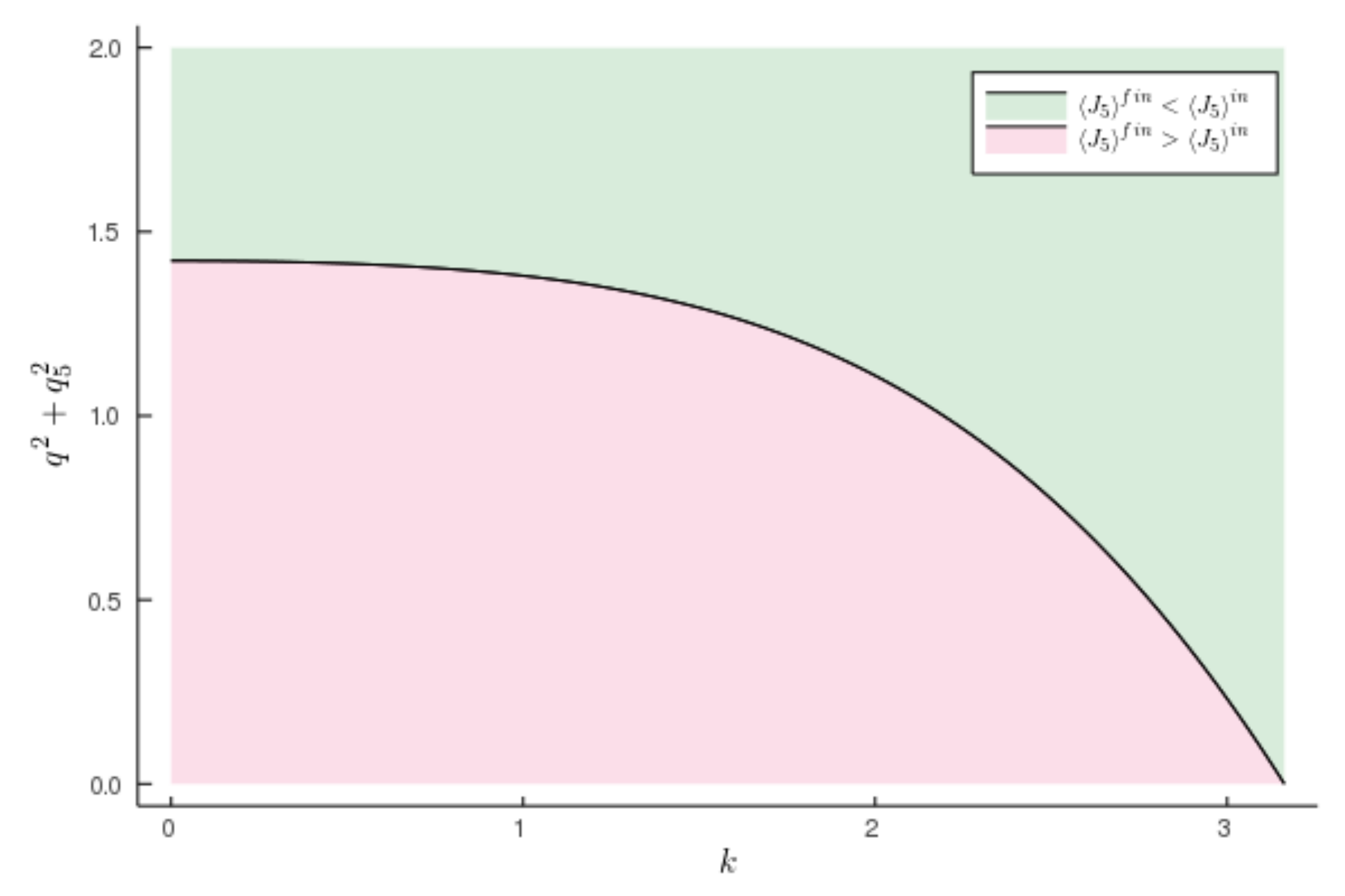}
    \caption{\small{Regions of the parameter space which lead to higher or lower final value of the axial current as compared to its initial value. The black line correspond to the limit where both values are equal. The upper half region give a final axial current below the initial value. Accordingly, in the lower half region one finds the final axial current above its initial value.}}
    \label{fig:4}
\end{figure}
\begin{figure}[h!]
    \centering
    \includegraphics[scale=0.55]{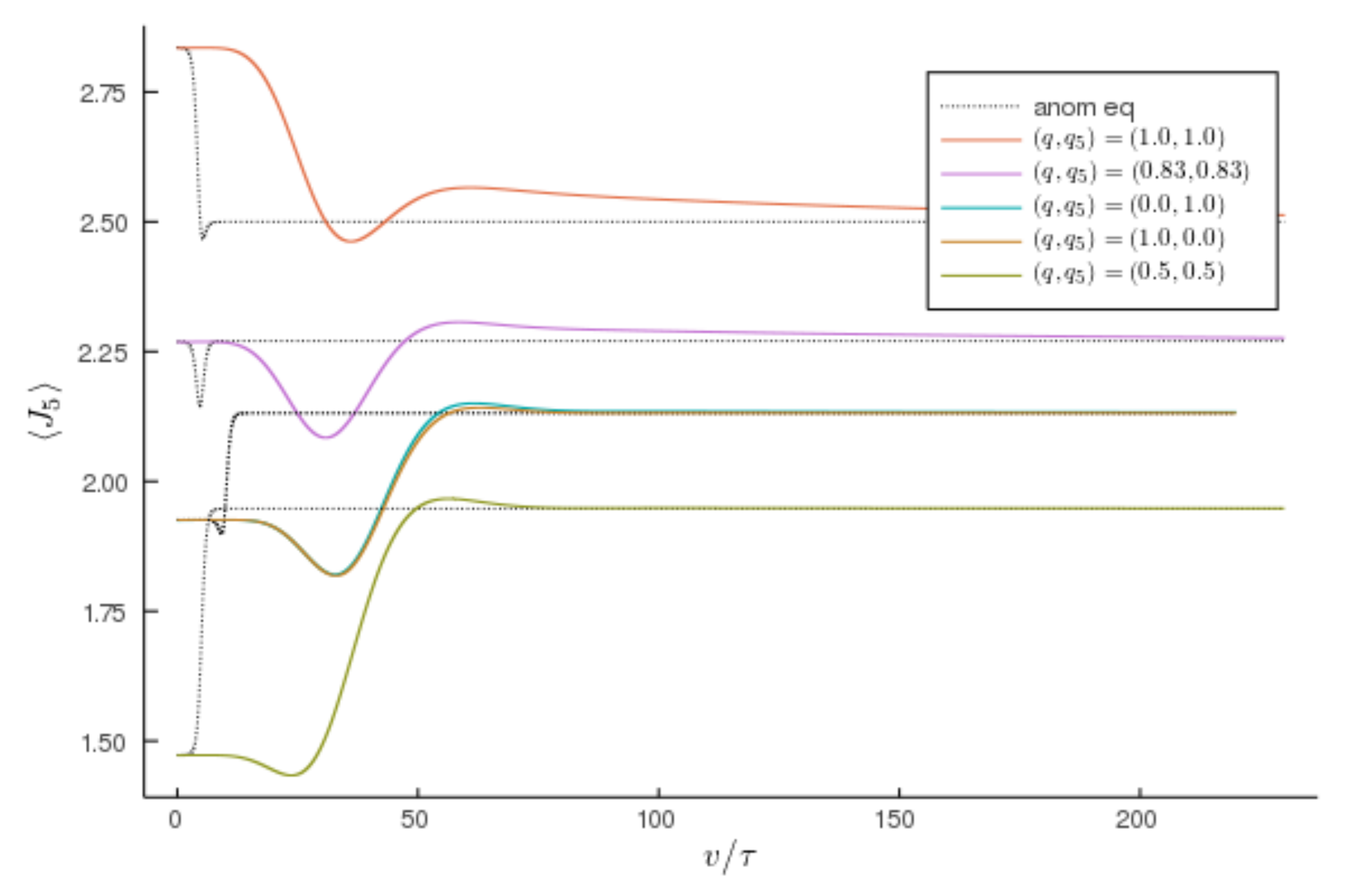}
    \caption{Axial current for various charges. We fix $\tau=0.05$ and $k=1.0$. We pick values on the parameter space for the three different regions of figure \ref{fig:4}, where the current globally diminishes, increases or maintains its initial value.}
    \label{fig:-7}
\end{figure}

As we discussed previously, there is in general an interplay between an increasing temperature and a decreasing chemical potential which is particularly relevant for the axial current. Already from equilibrium \ref{eq:3.5} we observe that one can have a final state with less or more current than the original one depending on the relative strength of both terms involved, which is determined by the point we are sitting on the parameter space. In figure \ref{fig:4} we show the regions on the parameter space that lead to each behavior. The black limiting line corresponds to configurations in which the initial and final currents are the same:

\begin{equation}
\label{eq:IV.2}
\begin{split}
    \left<J_5^z\right>_{in}&=  (q^2+q_5^2) + \dfrac{1}{12} \left(4-\dfrac{k^2}{2}-\dfrac{1}{6}(q^2+q_5^2)\right)^2\\
    &= \dfrac{256}{625}(q^2+q_5^2)+\dfrac{1}{12}\left(5-\dfrac{2k^2}{5}-\dfrac{512}{9375}(q^2+q_5^2)\right)^2 = \left<J_5^z\right>_{end}\, ,
\end{split}
\end{equation}


\noindent
where we have already substituted the values of $\alpha,\,\lambda,\,B_g,\,\mu_{(5)}\, \textrm{and} \,T$ at the initial and final states. The previous equation is not valid for $k=0$, as one then needs to include the convective flow.

\begin{figure}[h!]
    \centering
    \includegraphics[scale=0.48]{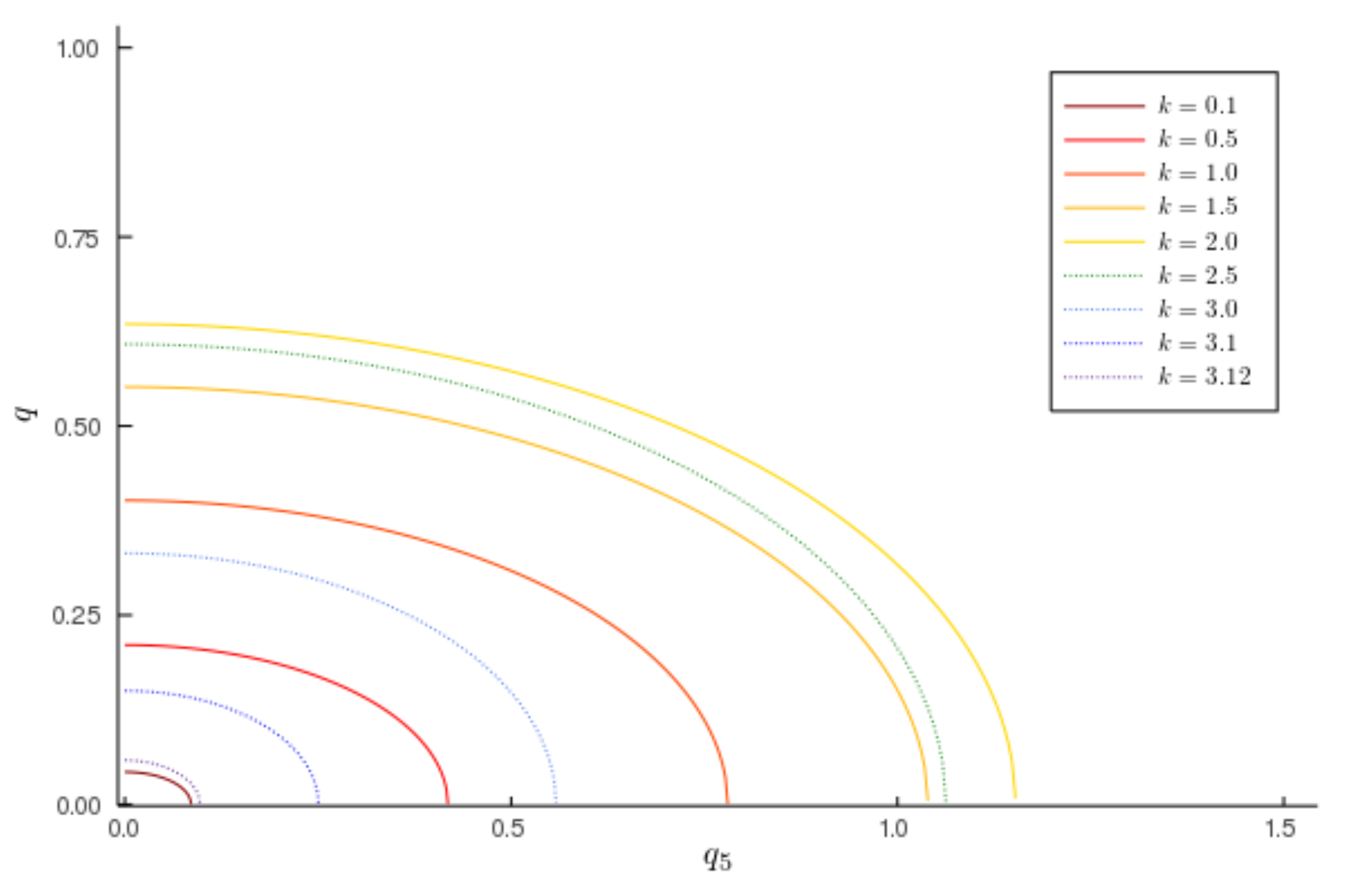}
    \caption{\small{Regions of the parameter space which lead to higher or lower final value of the energy current as compared to its initial value. Contour lines correspond to the limiting regions for fixed $k$. The inner regions produce enhancement of the energy current, whereas the outer regions lead to decrease of the energy current.}}
    \label{fig:4b}
\end{figure}

Sitting on the upper (green) half of the plane one finds a final axial current below its initial value. Accordingly, the lower (pinkish) region will give a final axial current above its initial value. Similarly, equation \ref{eq:3.6} also enjoys the interplay between the axial and gravitational anomaly. We can play the same game as before to find configurations where the final state has more energy current than its initial state. The regions of the parameter space displaying each behavior are shown in figure \ref{fig:4b}.  

\newpage

\subsection{The Role of the Gravitational Anomaly}

We have already seen that the inclusion of the gravitational anomaly gives a much richer structure, especially for the axial current. Even though in practice one has to deal with all anomalies at once, theoretically one can study the effect of each anomaly separately simply setting the remaining topological couplings to zero. The form of the equations of motion at linear order further implies that the currents can be constructed as the sum of the solutions only with $\alpha$ and only with $\lambda$. In the first case the equations of motion show that both axial and vector currents display similar behaviors, and no new interesting features are observed. Thus we focus in the situation where only the mixed gravitational anomaly is active. From the equilibrium predictions we know that the axial current can only increase, for the temperature is also increasing. On the contrary, the vector current vanishes identically in equilibrium. Any non-trivial response in the vector current must be interpreted as a sign of collective flow of the medium. The equations of motion reveal that it indeed couples to the energy current which is sensitive to the gravitational anomaly.
We study the effect of the gravitational anomaly for different values of the quench time and for different values of the charges.

\begin{figure}[h!]
    \centering
    \subfloat{\includegraphics[scale=0.5]{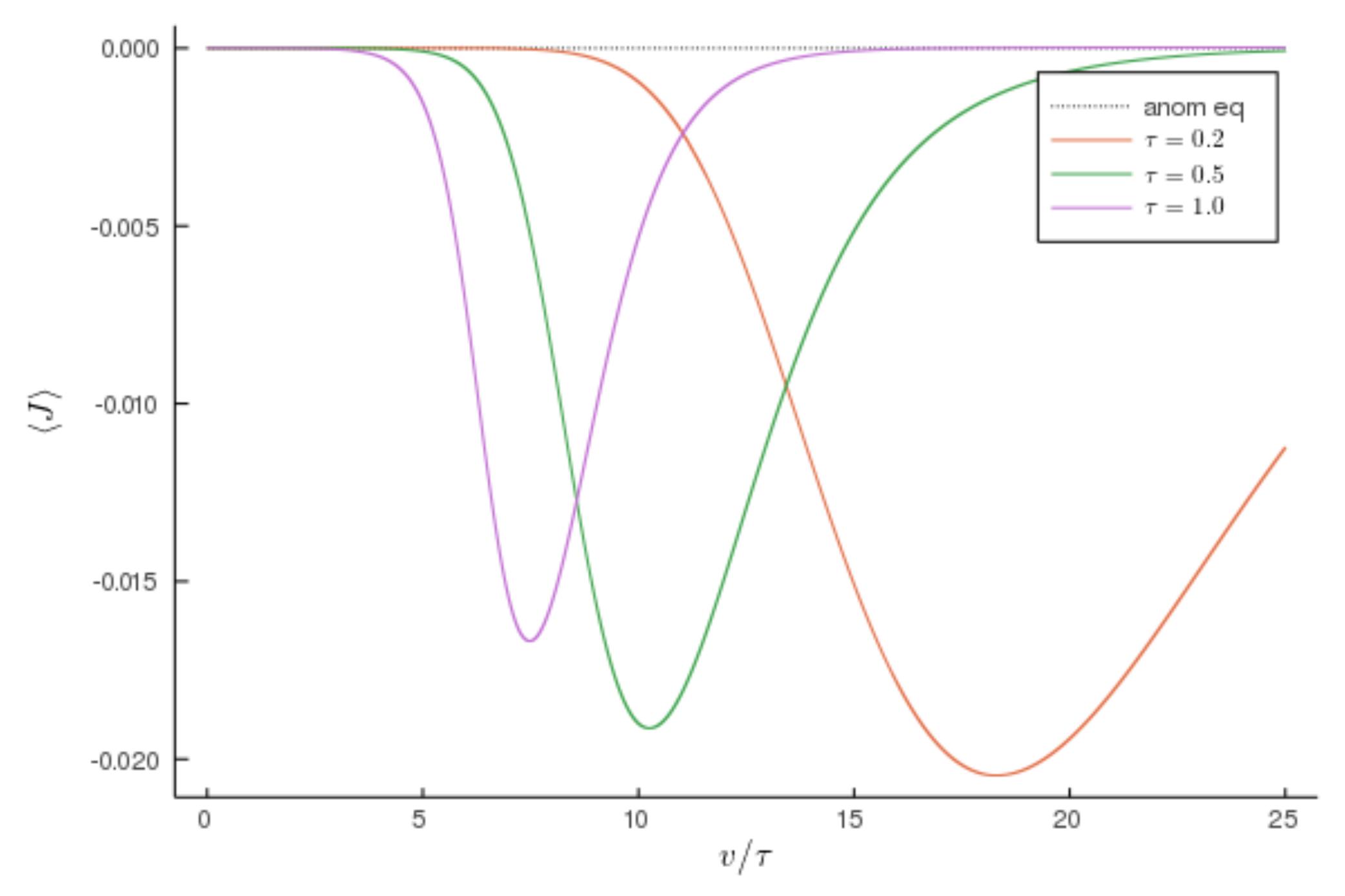}}%
    \qquad
    \subfloat{\includegraphics[scale=0.5]{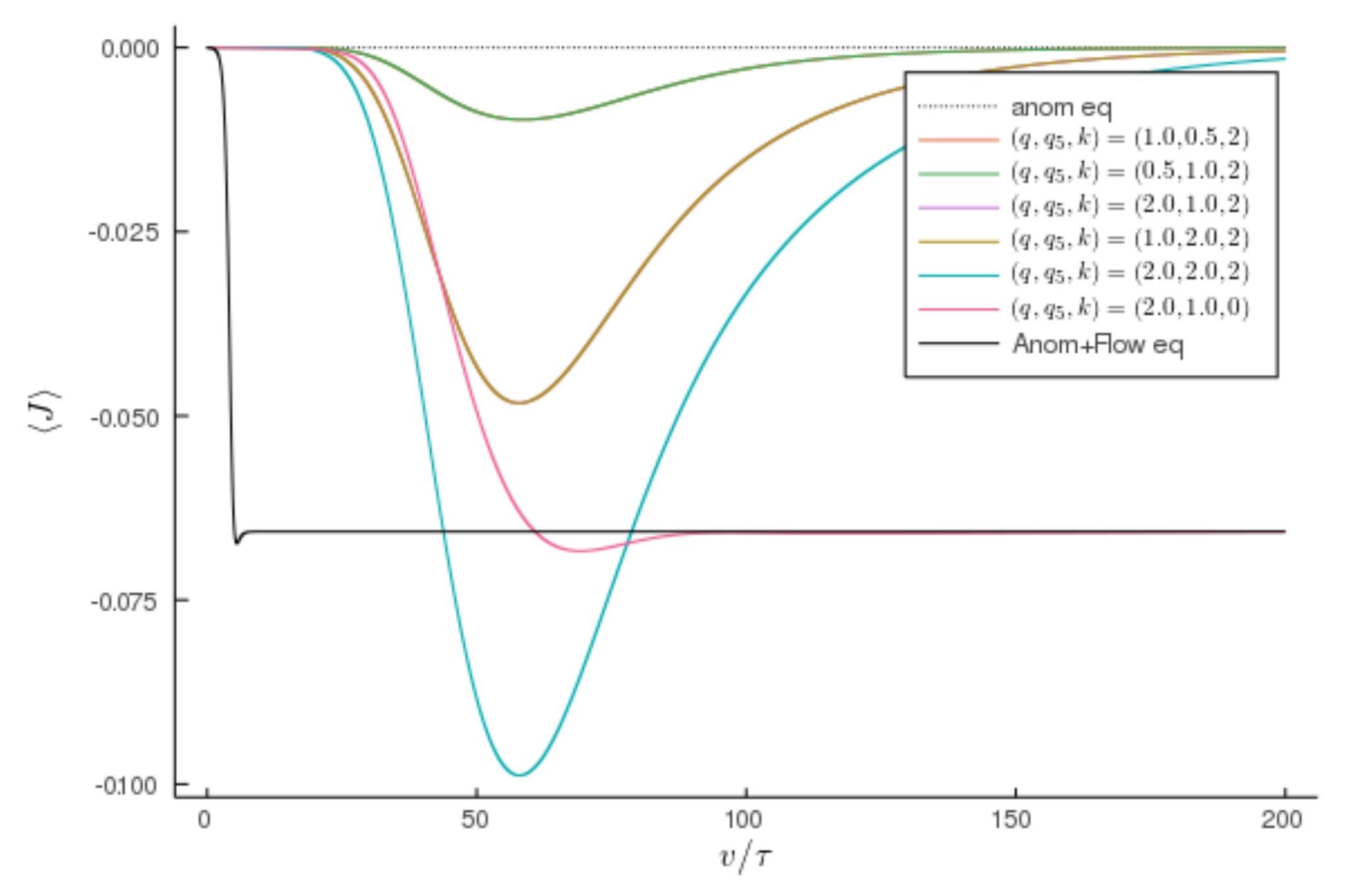}}
    \caption{\small{Vector current arising from the gravitational anomaly. We fix both charges to unity in the upper plot, $\tau=0.05$ for the lower plot, and $k=2.0$ for both of them. In the lower figure we have also included one case with $k=0$, which shows that the vector current response to the gravitational anomaly is mainly driven by the convective flow. As predicted, a vector current builds up even though the purely anomalous equilibrium prediction is that it should vanish at all times. The magnitude of this effect is very sensitive to the values of the charges. It is also affected by the value of the quench parameter in the usual manner, this is, the faster the quench the more it deviates from equilibrium. When momentum is conserved we end up with a non-trivial value for the vector current.}}
    \label{fig:-9}
\end{figure}

In figure \ref{fig:-9} we show the response of the vector current due to the gravitational anomaly. Even though the purely anomalous contribution vanishes identically, the dissipative convective flow is expected to participate, giving rise to a non-trivial vector current. The upper plot shows dependence with the quench parameter. The vector current gains relevance for faster quenches. As the full response is the sum of both anomalies separately, we can compare with figure \ref{fig:-3}. The contribution from the gravitational anomaly is then two orders of magnitude below the total value.  
The lower plot of figure \ref{fig:-9} shows dependence with both charges. In all of them we fix $k=2$ except for one of the currents, where we fix $k=0$. The momentum conserving case allows us to explicitly see the contribution of the convective flow to the vector current. As a consequence we end up with a non-trivial value for this current. 
Notice also that the vector current is symmetric under the exchange $q \leftrightarrow q_5$. In equilibrium this is a trivial statement, whereas out of equilibrium the situation is more delicate. The equation of motion driving the dynamics of the vector current \ref{eq:2.32} manifestly shows this symmetry except for the $qh'$ contribution. We should take a closer look at the dynamical equation for $h$ \ref{eq:2.34}. The constraint \ref{eq:consraint} allows to replace $qdv_z+q_5da_z$ in \ref{eq:2.34}\,. Along with the dynamical equation for $Z$ \ref{eq:2.31} this leaves us with a system of equations for $h$ and $Z$ whose source is proportional to $q_5$. Thus, we can rescale both fields as $h=q_5 \tilde{h}$ and $Z = q_5 \tilde Z$ so that the prefactor $q_5$ cancels out and the equation of motion for the tilde functions are now symmetric under the exchange of charges. Therefore $qh'=qq_5\tilde{h}'$ and the symmetry in the vector current is manifest. The previous discussion remains valid even when the axial anomaly is active ($\alpha \neq 0$). The reason why the symmetry does not show up with axial anomaly is that the initial equilibrium configuration breaks it (see equation \ref{eq:3.6}).   

\begin{figure}[h!]
    \centering
    \subfloat{\includegraphics[scale=0.5]{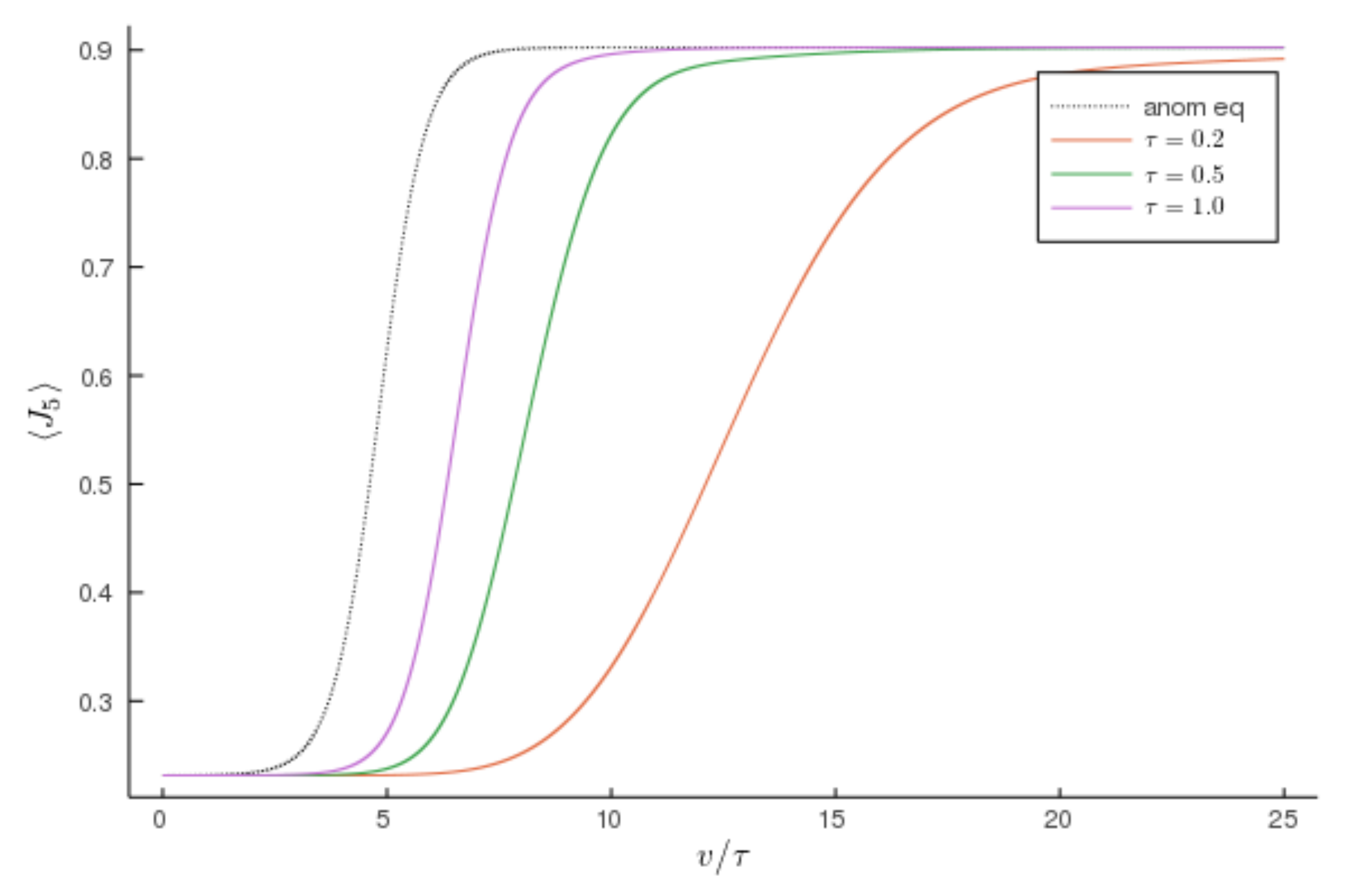}}%
    \qquad
    \subfloat{\includegraphics[scale=0.5]{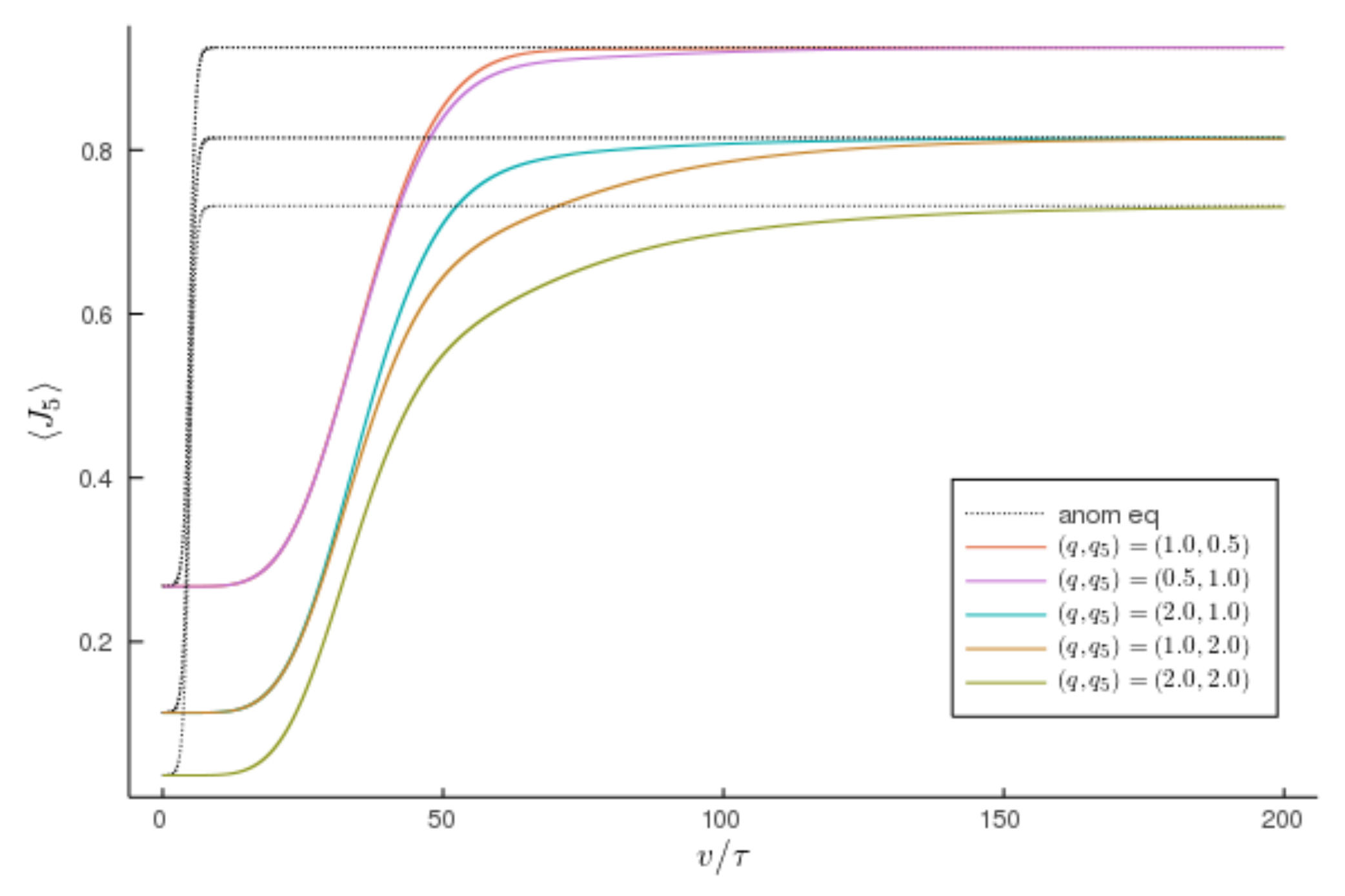}}
    \caption{\small{Axial current arising from the gravitational anomaly. We fix both charges to unity in the upper plot, $\tau=0.05$ for the lower plot, and $k=2.0$ for both of them. As temperature is increasing in our construction, so does the axial current. We observe that even though the equilibrium current is symmetric under $q \leftrightarrow q_5$, out of equilibrium this is no longer true. Bigger axial charge results in a behavior that departs more significantly from equilibrium. }}
    \label{fig:-10}
\end{figure}

\begin{figure}[h!]
    \centering
    \subfloat{\includegraphics[scale=0.5]{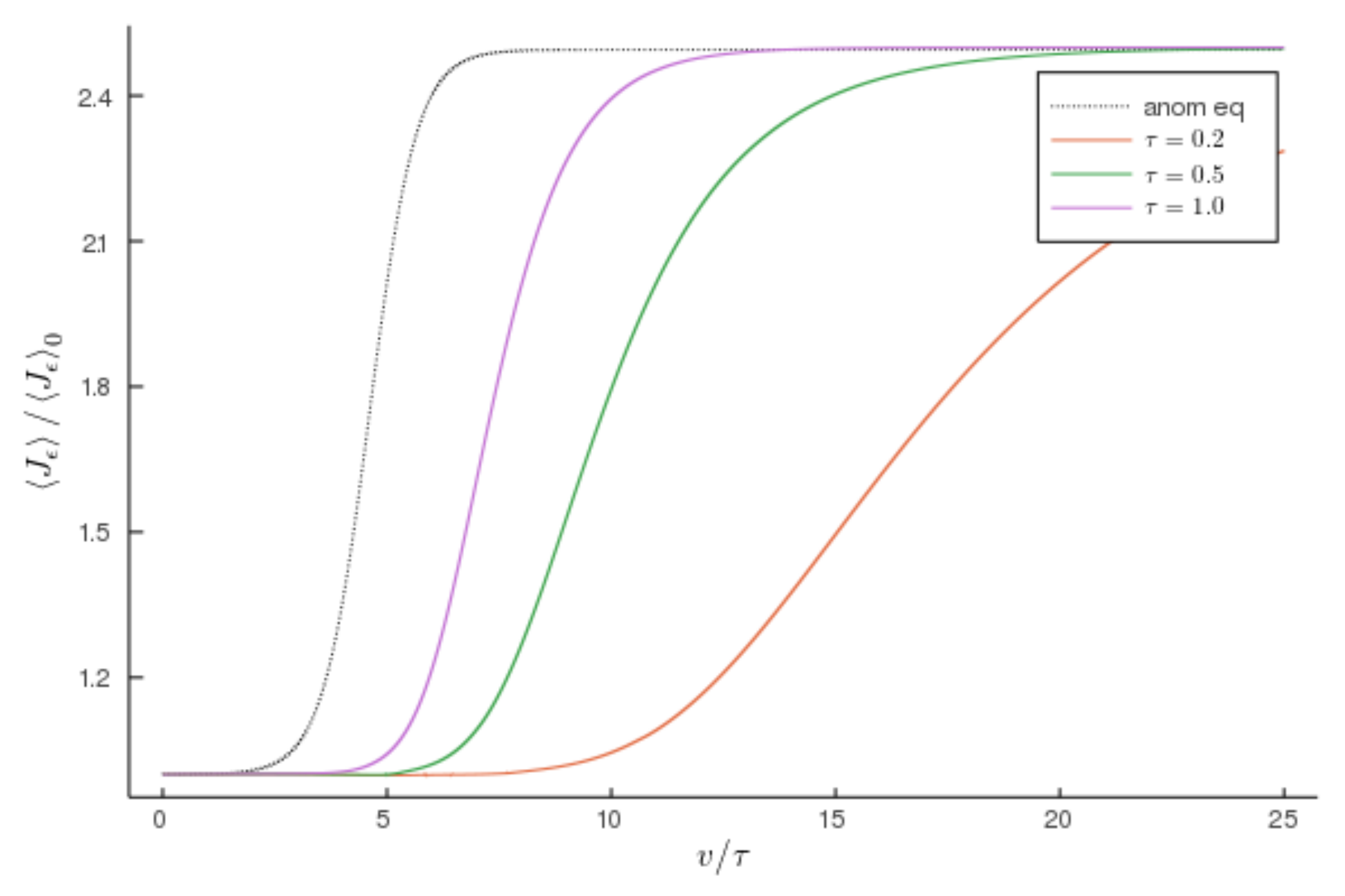}}%
    \qquad
    \subfloat{\includegraphics[scale=0.5]{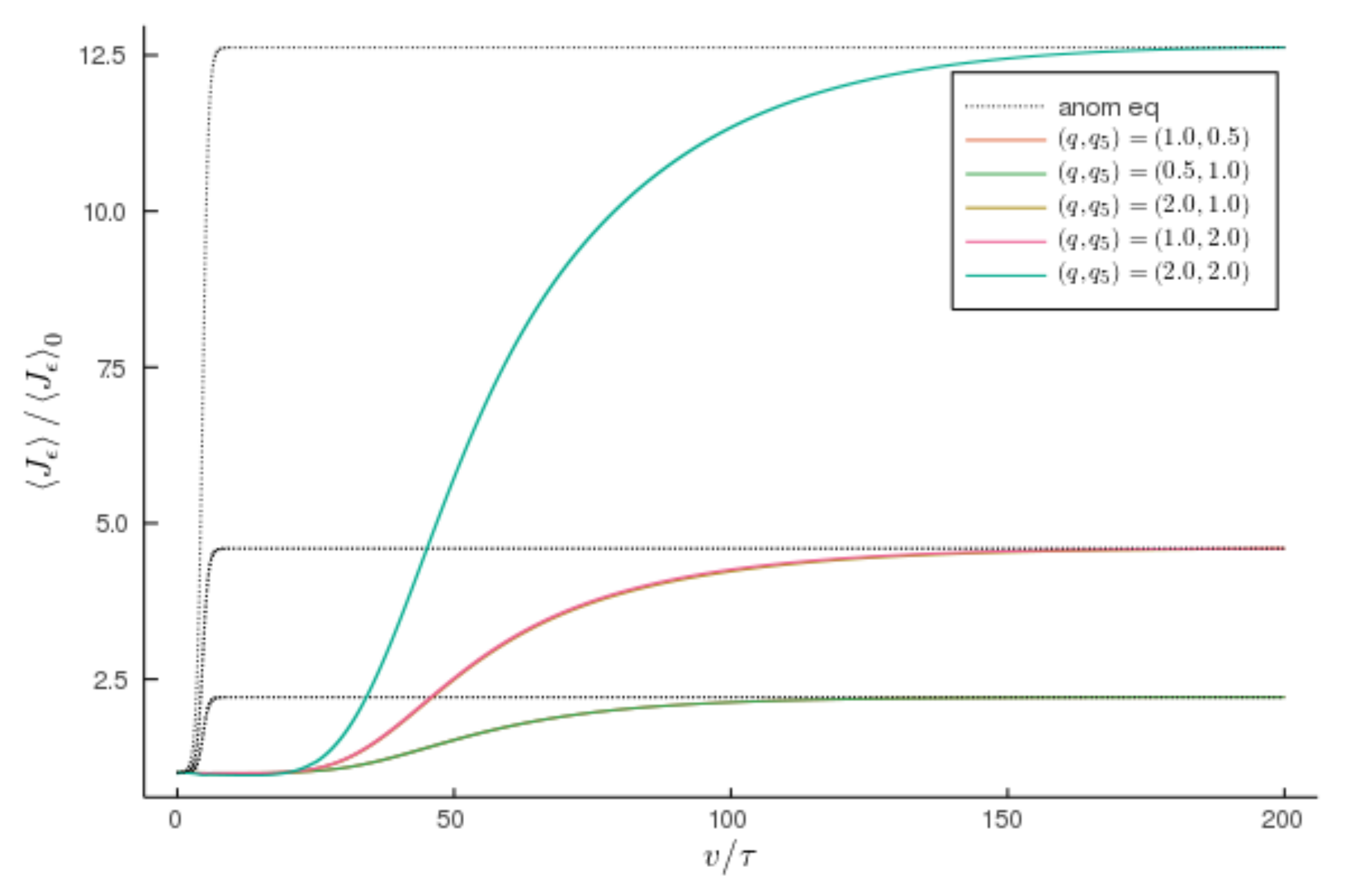}}
    \caption{\small{Energy current arising from the gravitational anomaly normalised to its initial value. We fix both charges to unity in the upper plot, $\tau=0.05$ for the lower plot, and $k=2.0$ for both of them. The effect of the gravitational anomaly is to increase the energy current. The normalised current is also symmetric under $q \leftrightarrow q_5$ in and out of equilibrium, even though this is not true for the un-normalized one.}}
    \label{fig:-11}
\end{figure}

In figures \ref{fig:-10} and \ref{fig:-11} we show the corresponding results for the axial and energy currents respectively. The axial ``gravitational" current is again symmetric under $q \leftrightarrow q_5$ but only in equilibrium! Out of it, the deviation from equilibrium is more significant for configurations with $q_5>q$. Finally, the ``gravitational" energy current is also symmetric under the interchange of charges in and out of equilibrium but only once one has divided by its initial value. 

\newpage

\newpage

\newpage

\newpage

\subsection{Response delay}

Another feature that the chiral vortical effect shares with the chiral magnetic effect in this construction is the delay for the onset of response after the quench. We follow the same criteria of \cite{Fernandez-Pendas:2019rkh} for the sake of comparison and compute the delay for the vector current. 
In particular, the quench is considered to be finished when the value of $4 \alpha \mu \mu_5 B_g$ with $\mu_{(5)}$ as in \ref{eq:II.14} deviates less than $0.1\%$ from the final equilibrium value. The time at which the quench finishes is denoted by $t_{quench}$\, and can be computed analytically. Analogously, the build-up in the current is considered to start when its value deviates more than $0.1\%$ from the initial value. We denote this time $t_{build\,up}$. The delay is defined thus as the difference between those two instants: $\Delta=t_{build\,up}-t_{quench}$\,.
Here we also expect the delay to depend mainly on the momentum relaxation $k$ and the quench parameter $\tau$.

\begin{figure}[h]
    \centering
    \includegraphics[scale=0.5]{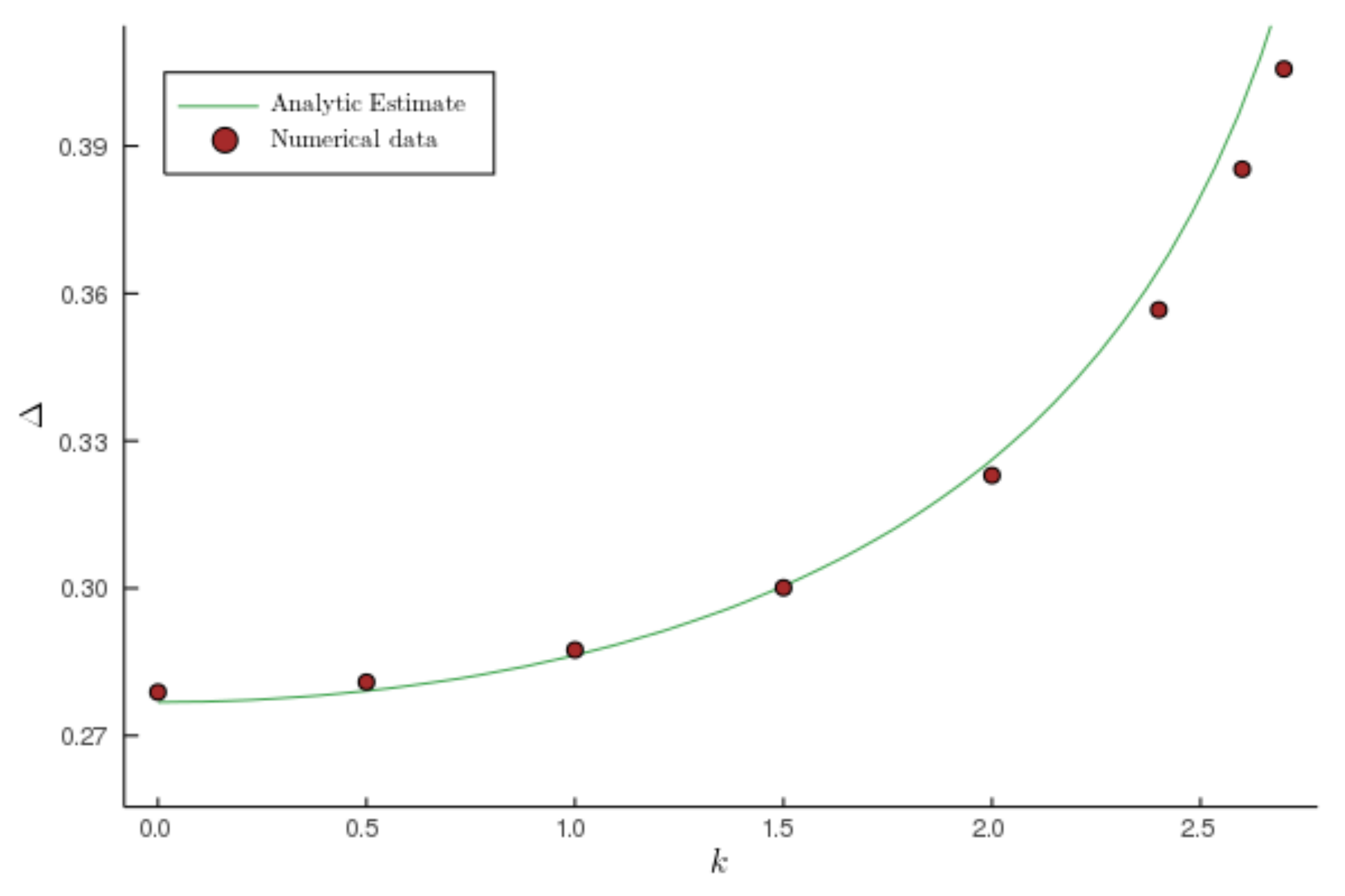}
    \caption{\small{Time delay as a function of the momentum relaxing parameter $k$ for fixed $\tau=2\cdot 10^{-4}\,$. In the analytical estimate, the integration point for each $k$ has been chosen so that the relative change in the blackening factor is $\delta f/f=0.11\,$. The estimation deviates for increasing $k$.}}
    \label{fig:5a}
\end{figure}

\begin{figure}[h]
    \centering
    \includegraphics[scale=0.5]{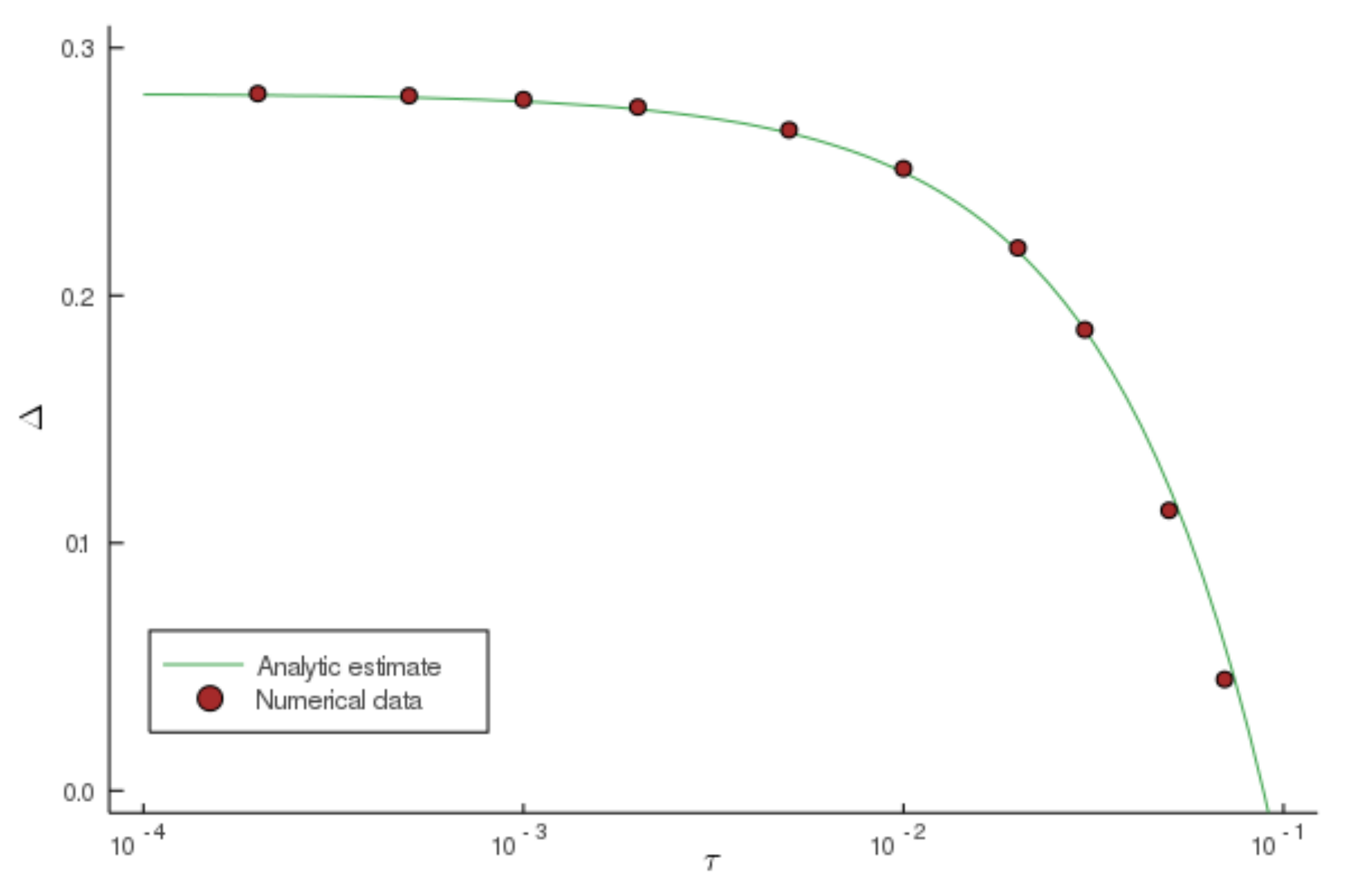}
    \caption{\small{Time delay as a function of the quench parameter $\tau$ for fixed $k=0\,$. The ``instantaneous quench approximation" we are using seems enough to reach agreement between the analytical estimate and the numerical data. For higher values of $\tau$ the approximation ceases to work satisfactorily and one needs to include the effects of the smooth quench. }}
    \label{fig:5b}
\end{figure}

The results obtained are shown in figures \ref{fig:5a} and \ref{fig:5b}. Not very surprisingly, the result is qualitatively very similar to that for the CME obtained in \cite{Fernandez-Pendas:2019rkh}. 
We observe that $\Delta$ approaches a finite well defined value for $\tau \to 0$\,. For high enough values of $\tau$ we can find that the delay becomes negative. This is an artifact of our definition and simply means that the current starts to build before the quench has finished. While the overall behaviour is very similar for CVE and CME the physical consequences in heavy ion collisions might be quite different. The magnetic field is generally believed to be very short lived. On the contrary vorticity is an intrinsic property of the quark gluon plasma and therefore is likely to be present during all of its lifetime. 

\subsubsection{Theoretical estimate}

A natural question to ask is whether (and how) one can extract the time delay in the build up of the current directly from the equations of motion, or more generically, from the model we have set up. To understand the origin of the delay we shall assume the quench is instantaneous ($t_{quench}=0$). We are quenching the mass of the black brane, or equivalently the blackening factor of the metric. Obviously, this is an object that lives in the bulk, whereas the (holographic) currents naturally live on the boundary. Besides, as the mass enters the blackening factor through $m u^2$, the boundary ($u\to 0$) cannot feel the change instantaneously. Thus, if we change abruptly $m$ at some time $v_q$, the current should not immediately change. The change in $m$ is however relevant in the bulk, so the question is how much time does it take for this information to reach the boundary. Intuitively this is what gives $t_{build\,up}$. In AdS, only massless particles can get to the boundary in a finite time $v_b$, and so the relevant piece of information should travel along a null geodesic. This allows us to find $v_b$ as a function of the depth $u$. Specifically, 

\begin{equation}
\label{eq:4.2b}
    v_b(u) = - \int_u^0\dfrac{2}{u_*^2 f(u_*)}  du_*,
\end{equation}

\noindent
where we have assumed that the quench is instantaneous and that we are only interested in $v_b$ after the quench has finished, so that the blackening factor remains constant in time.

The definition of the delay is somewhat arbitrary. This arbitrariness translates into the choice of $u$ in \ref{eq:4.2b}. We can state that when a photon located at some $u_i$ reaches the boundary, enough information has arrived to produce the desired change in the current. The point $u_i$ should not be taken near the boundary because the change in $f(v,u)$ is not significant, and nor should it be near the horizon, as by the time the signal arrives at the boundary the current must have felt the change already. We can pick $u_i$ so that the delay matches that of figure \ref{fig:5b} for $k=0$, $\tau \to 0\,$. This would give $u_i\simeq 0.53\,$. The relative change produced in the blackening factor is then 

\begin{equation}
    \dfrac{\delta f}{f} \equiv \left|\dfrac{f_{fin}-f_{in}}{f_{in}}\right|_{u_i\,,k}=\left|\dfrac{-2 (m_{fin}-m_{in}) u_i^2}{u_i^{-2} -\frac{1}{4}k^2-2m_{in}u_i^2 + \frac{1}{12}(q^2+q_5^2)u_i^4}\right| \simeq 0.11\,
\end{equation}

\noindent
where the subscripts $_{fin}$ and $_{in}$ denote the final and initial states respectively and we have evaluated the expression at $(u_i,k)=(0.53,0)\,$. We keep $\delta f/f$ fixed to $0.11$, as this value is the one producing the desired $0.1\%$ deviation in the current. Consequently the integration point $u_i$ must depend on $k$ to satisfy this constraint.

This is enough to find the $k$-dependence of the delay when the quench is instantaneous. The result is shown again in figure \ref{fig:5a}. The numerical agreement is remarkable, even though the estimation worsens for increasing $k$.

The dependence of the delay with $\tau$ is a bit trickier. As soon as $\tau \neq 0$ we must also have $t_{quench}\neq 0$. We compute $t_{quench}$ analytically from its definition. On the other hand $t_{build\,up}$ is also present. We compute it through \ref{eq:4.2b}, which means that we are neglecting the effect of a finite $\tau$ in $t_{build\,up}$\footnote{Including the effect of $\tau \neq 0$ in $t_{build\,up}$ reduces to solve the null geodesic equation for a time dependent blackening factor.}. As a consequence, our estimation is reliable for fast quenches only.  The result is shown in figure \ref{fig:5b}. Both numerical and analytical results show remarkable agreement in all the region where the delay stays positive, i.e. the current does not build up before the quench is finished.

\subsection{Quasinormal modes}
\label{sec:qnm}
The linear dependence of the equations of motion on the fields makes it reasonable to think that the temporal evolution may be well described in terms of the quasinormal modes, which are the intrinsic excitations of the system. However, the explicit time dependence on the blackening factor hinders this task, as the non-linear modes stemming from it should be included. A reasonably simpler question to ask is how fast a perturbation to the system will decay once we are at the final equilibrium state. In holography, the quasinormal modes are defined as the poles of the Green functions. We find them implementing numerically the determinant method \cite{Kaminski:2009dh}. In short, one replaces $\partial_v=-i\omega$ in the equations of motion with $B_g=0$ (because we are looking for intrinsic excitations). Then one finds linear independent solutions, arranges them in a matrix and solves for the frequencies that give a vanishing determinant on the boundary. We compute the QNM in the conventions relevant for the final state and we set $q=q_5=1.0$. For this choice of charges at $k=3.5$ the horizon becomes extremeness with vanishing  temperature. Therefore we stop our analysis there.  

\begin{figure}[h]
    \centering
    \includegraphics[scale=0.5]{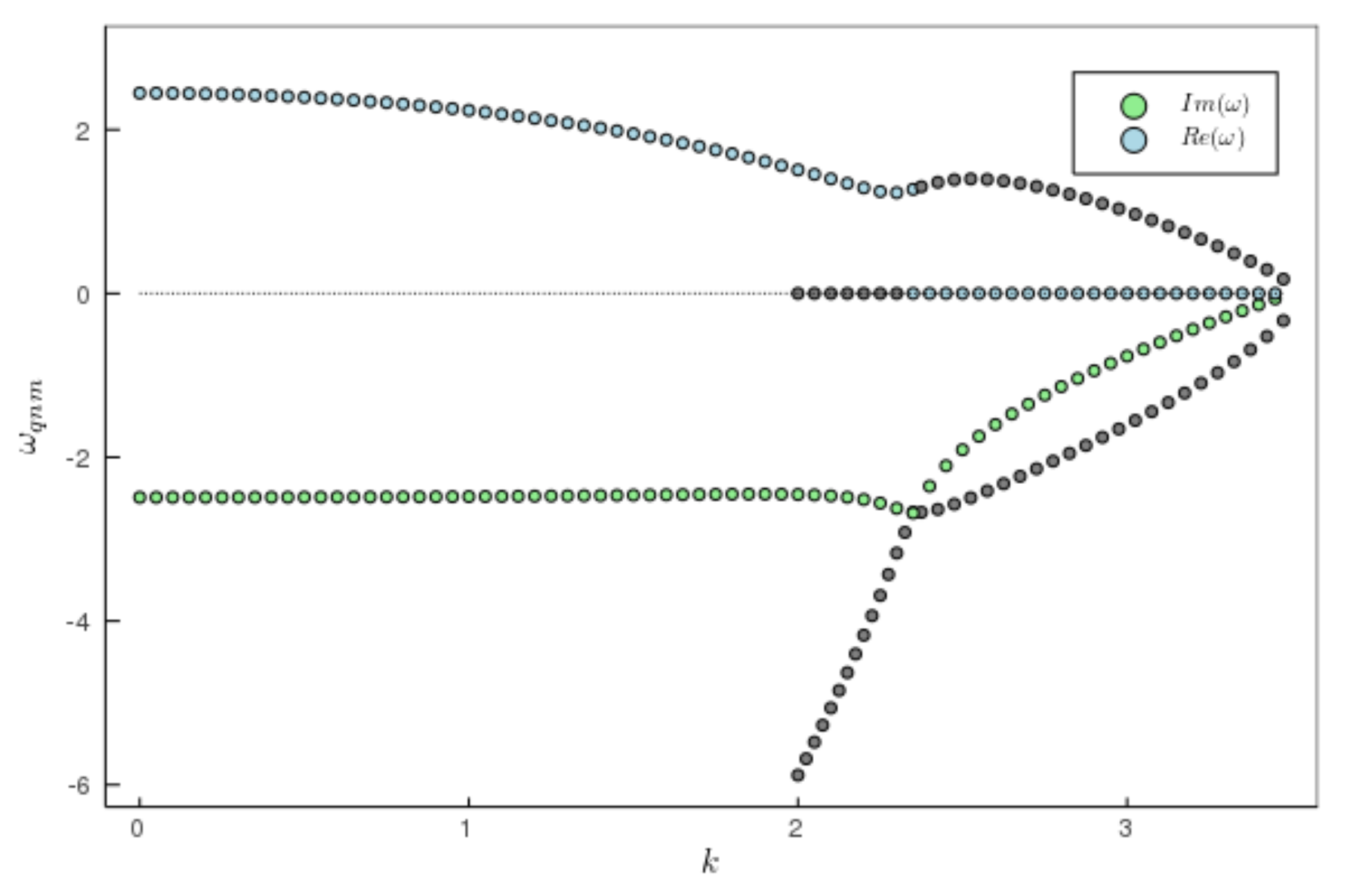}
    \caption{\small{Quasinormal modes of the field $\phi_1$ for different values of $k$. Around $k=2.3\,$, two branches of modes intersect one another. We keep the modes with lowest imaginary part in absolute value and shadow those which become less important.}}
    \label{fig:6}
\end{figure}

\begin{figure}[h]
    \centering
    \includegraphics[scale=0.5]{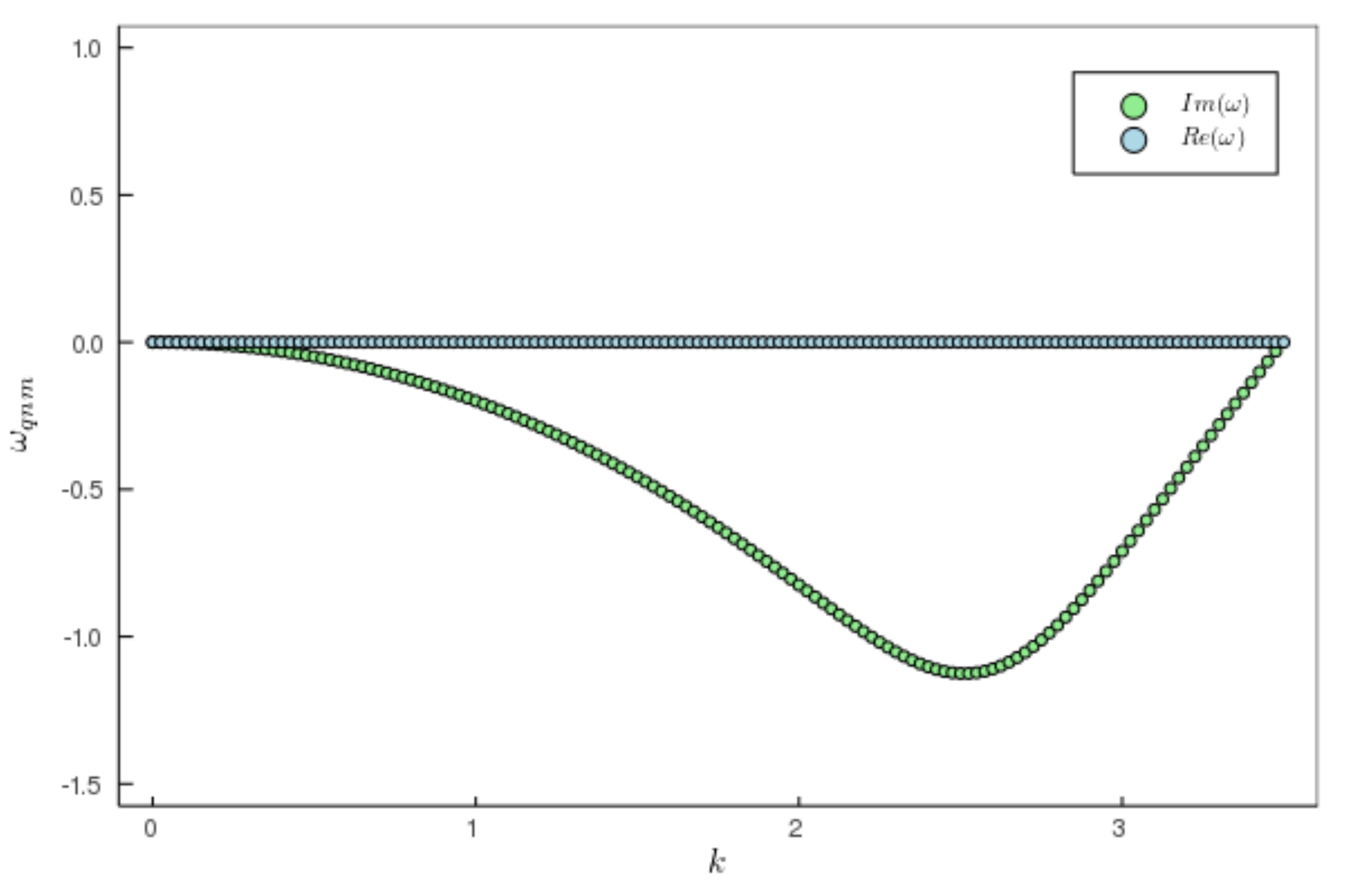}
    \caption{\small{Collective quasinormal modes -those defined for $\phi_2\,$, $Z$ and $h$- for different values of $k$. We observe that all modes have zero real part.
    The imaginary part reaches a minimum value at $k \simeq 2.5\,$, implying that the system relaxes faster for this particular $k$.}}
    \label{fig:7}
\end{figure}

\begin{figure}[h!]
    \centering
    \includegraphics[scale=0.46]{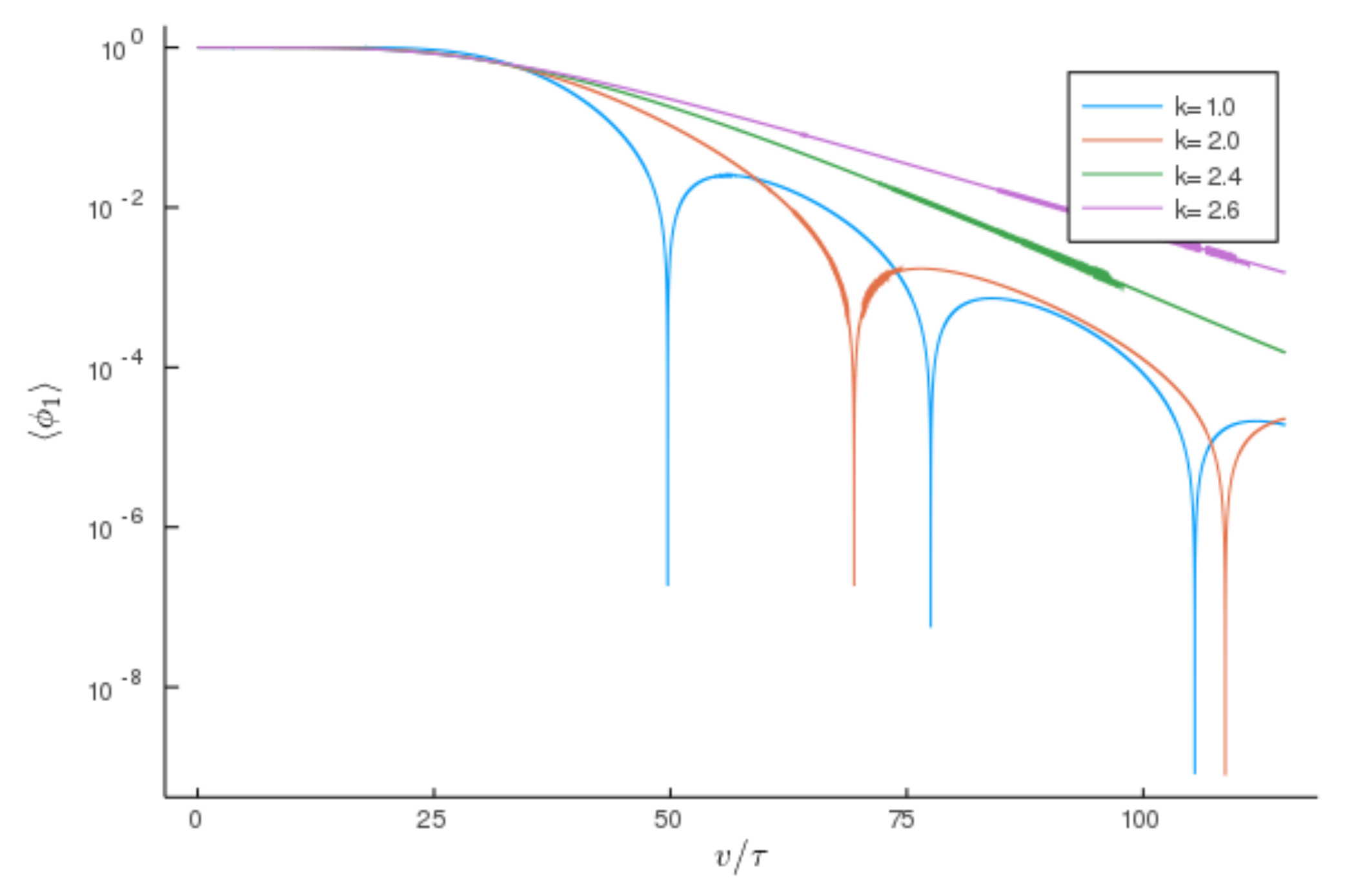}
    \caption{\small{Current associated to $\phi_1$ using the data from section \ref{sec:4A}. In agreement with the quasinormal modes in figure \ref{fig:6}, for $k \leq 2.3$ the currents show the characteristic ringdown related to the real part of the leading QNM. The slopes are related to the imaginary parts, which are the same for $k \leq 2.3$ and decreases thenceforth.}}
    \label{fig:log1}
\end{figure}

\begin{figure}[h]
    \centering
    \includegraphics[scale=0.45]{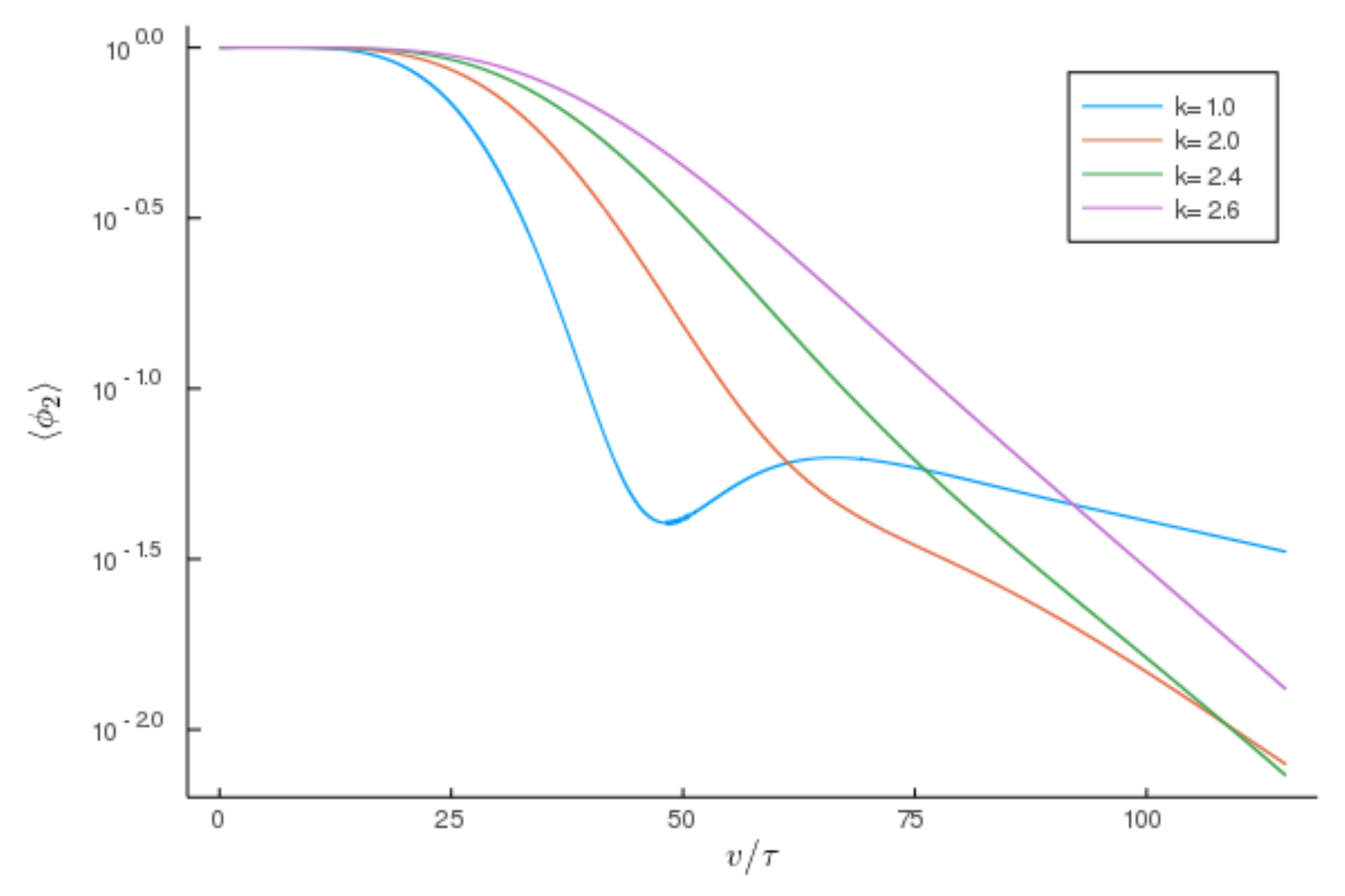}
    \caption{\small{Current associated to $\phi_2$ using the data from section \ref{sec:4A}. In agreement with the quasinormal modes in figure \ref{fig:7}, increasing the value of $k$ makes the current relax earlier up until $k=2.5$, after which the tendency is inverted. This is why the cases $k=2.4$ and $k=2.6$ have the same slope.}}
    \label{fig:log2}
\end{figure}

As the system of equations to solve is coupled, the quasinormal modes will be collective excitations, affecting all the fields the same in principle. However, in our particular case one can redefine the fields as 
\begin{equation}
\label{eq:4.2}
\begin{split}
    &\phi_1 = \dfrac{1}{\sqrt{q^2 + q_5^2}}(q a_z - q_5 v_z)\\
    &\phi_2 = \dfrac{1}{\sqrt{q^2 + q_5^2}}(q_5 a_z + q v_z ).
\end{split}
\end{equation}
%
The equations for the quasinormal modes are then

\begin{equation}
\begin{split}
   & u( u f\phi_1')'  - i\omega (\dfrac{\phi_1}{u}-2\phi_1')=0\,,\\
   & u( u f\phi_2')'  - i\omega (\dfrac{\phi_2}{u}-2\phi_2') - \sqrt{q^2+q_5^2} u h'=0\,,\\
   & u^3 ( \dfrac{1}{u}f Z')' -i \omega (3\dfrac{Z}{u}-2Z') -3k\dfrac{h}{u} + k h' = 0\,,\\
   & h'' - 3\dfrac{h'}{u} + k Z'-u^3 \sqrt{q^2+q_5^2} \phi_2 = 0\,.
\end{split}
\end{equation}
\noindent
Clearly $\phi_1$ decouples from the system, so independent quasinormal modes can be defined for it. They are shown in figure \ref{fig:6}. Around $k \simeq 2.3$ two branches of modes collide and we keep track only of the long-lived ones. The other field $\phi_2$ stays coupled and an effective charge $q_{eff}=\sqrt{q^2 + q_5^2}$ can be defined for it. The collective quasinormal modes are depicted in figure \ref{fig:7}. We have focused only in the modes with lowest imaginary time in absolute value, which are the ones governing the late time physics. Now the result is completely analogous to the excitations found for the CME. As a matter of fact, under the redefinitions \ref{eq:4.2}, the system of equations for the quasinormal modes in the CME and CVE coincide. Therefore the quasinormal modes spectrum is essentially the same as in \cite{Fernandez-Pendas:2019rkh} up to a slightly different value of the effective charge. In figures \ref{fig:log1} and \ref{fig:log2} we use the same data of section \ref{sec:4A} and plot them in a logarithmic scale so that the description in terms of quasinormal modes becomes more evident. 

The axial and vector currents get contributions from both collective and decoupled quasinormal modes according to \ref{eq:4.2}. Then the transition between the two regimes in figures \ref{fig:1} and \ref{fig:2} can be understood as the decoupled mode (figure \ref{fig:6}) loosing its real part at around $k\simeq2.3\,$. The relaxation times also agree with the collective quasinormal modes.

\newpage

\section{Conclusion and Outlook}
We have studied the far from equilibrium behavior of the CVE in a strongly coupled holographic quantum field theory. While similar studies have been reported before in the case of the CME this is the first one that concentrates on the CVE. 

One distinguishing feature of the equilibrium CVE is the temperature dependent term in the axial current. In our holographic model it is a consequence of the presence of a mixed gauge-gravitational Chern-Simons term, the holographic implementation of the gravitational contribution to the axial anomaly. This has important consequences. In our setup we keep the charges fixed and only introduce additional energy into the system leading to an increase in temperature and  a decrease in chemical potentials. Depending on the charges and momentum relaxation parameter the axial (and energy) currents can therefore either decrease or increase. 

Another important effect is that even in the absence of the purely gauge Chern-Simons term the vector current reacts to the presence of the gravitational anomaly. In cases with momentum conservation this can be understood as a convective transport due to the flow that is induced during the non-equilibrium evolution (similar to what has been observed in \cite{Landsteiner:2017lwm}.
With momentum relaxation the vector current reacts only during the non-equilibrium stages and returns to a trivial value once equilibrium is obtained. 

We have also observed a delay in the build-up of the CVE current. This is similar to what was found in for the CME \cite{Landsteiner:2017lwm,Fernandez-Pendas:2019rkh}. It can be taken as a sign that the CVE is significant only if a system is sufficiently close to (local) equilibrium. In the case of the CME this might have important consequences for the interpretation of experimental data in heavy ion collisions due to the short lifetime of the magnetic field. In the case of the CVE this is probably less relevant. In the quark gluon plasma vorticity is an intrinsic quantity and therefore will in general be present throughout the lifetime of the plasma. 

The picture that arises from this and previous studies on anomalous transport is that CVE (and CME) are effective only if the system is sufficiently close to equilibrium. A very important question is therefore if and how this can be quantified. Another important point is to go beyond the linear response regime. For the holographic models this means that the back reaction of the magnetic field or vorticity onto the metric has to be taken into account. We plan to investigate these issues in future work.

\acknowledgments{We thank J. Fernandez-Pend\'as for useful discussion and communication, M. Baggioli for discussion on the interpretation of the momentum relaxation out of equilibrium. This work was supported by Agencia Estatal de Investigaci{\'o}n  IFT
Centro de Excelencia Severo Ochoa SEV-2016-0597, and by the grant PGC2018-095976-B-C21 from MCIU/AEI/FEDER, UE. S.M.T. was supported by an FPI-UAM predoctoral fellowship. }

\bibliography{AnomTrans}

\begin{thebibliography}{37}%
\makeatletter
\providecommand \@ifxundefined [1]{%
 \@ifx{#1\undefined}
}%
\providecommand \@ifnum [1]{%
 \ifnum #1\expandafter \@firstoftwo
 \else \expandafter \@secondoftwo
 \fi
}%
\providecommand \@ifx [1]{%
 \ifx #1\expandafter \@firstoftwo
 \else \expandafter \@secondoftwo
 \fi
}%
\providecommand \natexlab [1]{#1}%
\providecommand \enquote  [1]{``#1''}%
\providecommand \bibnamefont  [1]{#1}%
\providecommand \bibfnamefont [1]{#1}%
\providecommand \citenamefont [1]{#1}%
\providecommand \href@noop [0]{\@secondoftwo}%
\providecommand \href [0]{\begingroup \@sanitize@url \@href}%
\providecommand \@href[1]{\@@startlink{#1}\@@href}%
\providecommand \@@href[1]{\endgroup#1\@@endlink}%
\providecommand \@sanitize@url [0]{\catcode `\\12\catcode `\$12\catcode
  `\&12\catcode `\#12\catcode `\^12\catcode `\_12\catcode `\%12\relax}%
\providecommand \@@startlink[1]{}%
\providecommand \@@endlink[0]{}%
\providecommand \url  [0]{\begingroup\@sanitize@url \@url }%
\providecommand \@url [1]{\endgroup\@href {#1}{\urlprefix }}%
\providecommand \urlprefix  [0]{URL }%
\providecommand \Eprint [0]{\href }%
\providecommand \doibase [0]{http://dx.doi.org/}%
\providecommand \selectlanguage [0]{\@gobble}%
\providecommand \bibinfo  [0]{\@secondoftwo}%
\providecommand \bibfield  [0]{\@secondoftwo}%
\providecommand \translation [1]{[#1]}%
\providecommand \BibitemOpen [0]{}%
\providecommand \bibitemStop [0]{}%
\providecommand \bibitemNoStop [0]{.\EOS\space}%
\providecommand \EOS [0]{\spacefactor3000\relax}%
\providecommand \BibitemShut  [1]{\csname bibitem#1\endcsname}%
\let\auto@bib@innerbib\@empty
\bibitem [{\citenamefont {Kharzeev}\ \emph {et~al.}(2016)\citenamefont
  {Kharzeev}, \citenamefont {Liao}, \citenamefont {Voloshin},\ and\
  \citenamefont {Wang}}]{Kharzeev:2015znc}%
  \BibitemOpen
  \bibfield  {author} {\bibinfo {author} {\bibfnamefont {D.~E.}\ \bibnamefont
  {Kharzeev}}, \bibinfo {author} {\bibfnamefont {J.}~\bibnamefont {Liao}},
  \bibinfo {author} {\bibfnamefont {S.~A.}\ \bibnamefont {Voloshin}}, \ and\
  \bibinfo {author} {\bibfnamefont {G.}~\bibnamefont {Wang}},\ }\href {\doibase
  10.1016/j.ppnp.2016.01.001} {\bibfield  {journal} {\bibinfo  {journal} {Prog.
  Part. Nucl. Phys.}\ }\textbf {\bibinfo {volume} {88}},\ \bibinfo {pages} {1}
  (\bibinfo {year} {2016})},\ \Eprint {http://arxiv.org/abs/1511.04050}
  {arXiv:1511.04050 [hep-ph]} \BibitemShut {NoStop}%
\bibitem [{\citenamefont {Landsteiner}(2016)}]{Landsteiner:2016led}%
  \BibitemOpen
  \bibfield  {author} {\bibinfo {author} {\bibfnamefont {K.}~\bibnamefont
  {Landsteiner}},\ }in\ \href
  {http://inspirehep.net/record/1492126/files/arXiv:1610.04413.pdf} {\emph
  {\bibinfo {booktitle} {{56th Cracow School of Theoretical Physics: A Panorama
  of Holography Zakopane, Poland, May 24-June 1, 2016}}}}\ (\bibinfo {year}
  {2016})\ \Eprint {http://arxiv.org/abs/1610.04413} {arXiv:1610.04413
  [hep-th]} \BibitemShut {NoStop}%
\bibitem [{\citenamefont {Kharzeev}\ and\ \citenamefont
  {Liao}(2019)}]{Kharzeev:2019zgg}%
  \BibitemOpen
  \bibfield  {author} {\bibinfo {author} {\bibfnamefont {D.}~\bibnamefont
  {Kharzeev}}\ and\ \bibinfo {author} {\bibfnamefont {J.}~\bibnamefont
  {Liao}},\ }\href {\doibase 10.1080/10619127.2018.1495479} {\bibfield
  {journal} {\bibinfo  {journal} {Nucl. Phys. News}\ }\textbf {\bibinfo
  {volume} {29}},\ \bibinfo {pages} {26} (\bibinfo {year} {2019})}\BibitemShut
  {NoStop}%
\bibitem [{\citenamefont {Adamczyk}\ \emph {et~al.}(2017)\citenamefont
  {Adamczyk} \emph {et~al.}}]{STAR:2017ckg}%
  \BibitemOpen
  \bibfield  {author} {\bibinfo {author} {\bibfnamefont {L.}~\bibnamefont
  {Adamczyk}} \emph {et~al.} (\bibinfo {collaboration} {STAR}),\ }\href
  {\doibase 10.1038/nature23004} {\bibfield  {journal} {\bibinfo  {journal}
  {Nature}\ }\textbf {\bibinfo {volume} {548}},\ \bibinfo {pages} {62}
  (\bibinfo {year} {2017})},\ \Eprint {http://arxiv.org/abs/1701.06657}
  {arXiv:1701.06657 [nucl-ex]} \BibitemShut {NoStop}%
\bibitem [{\citenamefont {Li}\ \emph {et~al.}(2016)\citenamefont {Li},
  \citenamefont {Kharzeev}, \citenamefont {Zhang}, \citenamefont {Huang},
  \citenamefont {Pletikosic}, \citenamefont {Fedorov}, \citenamefont {Zhong},
  \citenamefont {Schneeloch}, \citenamefont {Gu},\ and\ \citenamefont
  {Valla}}]{Li:2014bha}%
  \BibitemOpen
  \bibfield  {author} {\bibinfo {author} {\bibfnamefont {Q.}~\bibnamefont
  {Li}}, \bibinfo {author} {\bibfnamefont {D.~E.}\ \bibnamefont {Kharzeev}},
  \bibinfo {author} {\bibfnamefont {C.}~\bibnamefont {Zhang}}, \bibinfo
  {author} {\bibfnamefont {Y.}~\bibnamefont {Huang}}, \bibinfo {author}
  {\bibfnamefont {I.}~\bibnamefont {Pletikosic}}, \bibinfo {author}
  {\bibfnamefont {A.~V.}\ \bibnamefont {Fedorov}}, \bibinfo {author}
  {\bibfnamefont {R.~D.}\ \bibnamefont {Zhong}}, \bibinfo {author}
  {\bibfnamefont {J.~A.}\ \bibnamefont {Schneeloch}}, \bibinfo {author}
  {\bibfnamefont {G.~D.}\ \bibnamefont {Gu}}, \ and\ \bibinfo {author}
  {\bibfnamefont {T.}~\bibnamefont {Valla}},\ }\href {\doibase
  10.1038/nphys3648} {\bibfield  {journal} {\bibinfo  {journal} {Nature Phys.}\
  }\textbf {\bibinfo {volume} {12}},\ \bibinfo {pages} {550} (\bibinfo {year}
  {2016})},\ \Eprint {http://arxiv.org/abs/1412.6543} {arXiv:1412.6543
  [cond-mat.str-el]} \BibitemShut {NoStop}%
\bibitem [{\citenamefont {Gynther}\ \emph {et~al.}(2011)\citenamefont
  {Gynther}, \citenamefont {Landsteiner}, \citenamefont {Pena-Benitez},\ and\
  \citenamefont {Rebhan}}]{Gynther:2010ed}%
  \BibitemOpen
  \bibfield  {author} {\bibinfo {author} {\bibfnamefont {A.}~\bibnamefont
  {Gynther}}, \bibinfo {author} {\bibfnamefont {K.}~\bibnamefont
  {Landsteiner}}, \bibinfo {author} {\bibfnamefont {F.}~\bibnamefont
  {Pena-Benitez}}, \ and\ \bibinfo {author} {\bibfnamefont {A.}~\bibnamefont
  {Rebhan}},\ }\href {\doibase 10.1007/JHEP02(2011)110} {\bibfield  {journal}
  {\bibinfo  {journal} {JHEP}\ }\textbf {\bibinfo {volume} {02}},\ \bibinfo
  {pages} {110} (\bibinfo {year} {2011})},\ \Eprint
  {http://arxiv.org/abs/1005.2587} {arXiv:1005.2587 [hep-th]} \BibitemShut
  {NoStop}%
\bibitem [{\citenamefont {M.M.~Vazifeh}(2013)}]{FranzVazifeh}%
  \BibitemOpen
  \bibfield  {author} {\bibinfo {author} {\bibfnamefont {M.~F.}\ \bibnamefont
  {M.M.~Vazifeh}},\ }\href {\doibase 10.1103/PhysRevLett.111.027201} {\bibfield
   {journal} {\bibinfo  {journal} {Phys. Rev.Lett.}\ }\textbf {\bibinfo
  {volume} {111}},\ \bibinfo {pages} {027201} (\bibinfo {year} {2013})},\
  \Eprint {http://arxiv.org/abs/1303.5784} {arXiv:1303.5784 [cond-mat]}
  \BibitemShut {NoStop}%
\bibitem [{\citenamefont {Yamamoto}(2015)}]{Yamamoto:2015fxa}%
  \BibitemOpen
  \bibfield  {author} {\bibinfo {author} {\bibfnamefont {N.}~\bibnamefont
  {Yamamoto}},\ }\href {\doibase 10.1103/PhysRevD.92.085011} {\bibfield
  {journal} {\bibinfo  {journal} {Phys. Rev.}\ }\textbf {\bibinfo {volume}
  {D92}},\ \bibinfo {pages} {085011} (\bibinfo {year} {2015})},\ \Eprint
  {http://arxiv.org/abs/1502.01547} {arXiv:1502.01547 [cond-mat.mes-hall]}
  \BibitemShut {NoStop}%
\bibitem [{\citenamefont {Bardeen}(1969)}]{Bardeen:1969md}%
  \BibitemOpen
  \bibfield  {author} {\bibinfo {author} {\bibfnamefont {W.~A.}\ \bibnamefont
  {Bardeen}},\ }\href {\doibase 10.1103/PhysRev.184.1848} {\bibfield  {journal}
  {\bibinfo  {journal} {Phys. Rev.}\ }\textbf {\bibinfo {volume} {184}},\
  \bibinfo {pages} {1848} (\bibinfo {year} {1969})}\BibitemShut {NoStop}%
\bibitem [{\citenamefont {Vilenkin}(1980)}]{Vilenkin:1980zv}%
  \BibitemOpen
  \bibfield  {author} {\bibinfo {author} {\bibfnamefont {A.}~\bibnamefont
  {Vilenkin}},\ }\href {\doibase 10.1103/PhysRevD.21.2260} {\bibfield
  {journal} {\bibinfo  {journal} {Phys. Rev.}\ }\textbf {\bibinfo {volume}
  {D21}},\ \bibinfo {pages} {2260} (\bibinfo {year} {1980})}\BibitemShut
  {NoStop}%
\bibitem [{\citenamefont {Landsteiner}\ \emph
  {et~al.}(2011{\natexlab{a}})\citenamefont {Landsteiner}, \citenamefont
  {Megias},\ and\ \citenamefont {Pena-Benitez}}]{Landsteiner:2011tg}%
  \BibitemOpen
  \bibfield  {author} {\bibinfo {author} {\bibfnamefont {K.}~\bibnamefont
  {Landsteiner}}, \bibinfo {author} {\bibfnamefont {E.}~\bibnamefont {Megias}},
  \ and\ \bibinfo {author} {\bibfnamefont {F.}~\bibnamefont {Pena-Benitez}},\
  }in\ \href@noop {} {\emph {\bibinfo {booktitle} {{11th Workshop on
  Non-Perturbative Quantum Chromodynamics}}}}\ (\bibinfo {year} {2011})\
  \Eprint {http://arxiv.org/abs/1110.3615} {arXiv:1110.3615 [hep-ph]}
  \BibitemShut {NoStop}%
\bibitem [{\citenamefont {Amado}\ \emph {et~al.}(2011)\citenamefont {Amado},
  \citenamefont {Landsteiner},\ and\ \citenamefont
  {Pena-Benitez}}]{Amado:2011zx}%
  \BibitemOpen
  \bibfield  {author} {\bibinfo {author} {\bibfnamefont {I.}~\bibnamefont
  {Amado}}, \bibinfo {author} {\bibfnamefont {K.}~\bibnamefont {Landsteiner}},
  \ and\ \bibinfo {author} {\bibfnamefont {F.}~\bibnamefont {Pena-Benitez}},\
  }\href {\doibase 10.1007/JHEP05(2011)081} {\bibfield  {journal} {\bibinfo
  {journal} {JHEP}\ }\textbf {\bibinfo {volume} {05}},\ \bibinfo {pages} {081}
  (\bibinfo {year} {2011})},\ \Eprint {http://arxiv.org/abs/1102.4577}
  {arXiv:1102.4577 [hep-th]} \BibitemShut {NoStop}%
\bibitem [{\citenamefont {Erdmenger}\ \emph {et~al.}(2009)\citenamefont
  {Erdmenger}, \citenamefont {Haack}, \citenamefont {Kaminski},\ and\
  \citenamefont {Yarom}}]{Erdmenger:2008rm}%
  \BibitemOpen
  \bibfield  {author} {\bibinfo {author} {\bibfnamefont {J.}~\bibnamefont
  {Erdmenger}}, \bibinfo {author} {\bibfnamefont {M.}~\bibnamefont {Haack}},
  \bibinfo {author} {\bibfnamefont {M.}~\bibnamefont {Kaminski}}, \ and\
  \bibinfo {author} {\bibfnamefont {A.}~\bibnamefont {Yarom}},\ }\href
  {\doibase 10.1088/1126-6708/2009/01/055} {\bibfield  {journal} {\bibinfo
  {journal} {JHEP}\ }\textbf {\bibinfo {volume} {01}},\ \bibinfo {pages} {055}
  (\bibinfo {year} {2009})},\ \Eprint {http://arxiv.org/abs/0809.2488}
  {arXiv:0809.2488 [hep-th]} \BibitemShut {NoStop}%
\bibitem [{\citenamefont {Banerjee}\ \emph {et~al.}(2011)\citenamefont
  {Banerjee}, \citenamefont {Bhattacharya}, \citenamefont {Bhattacharyya},
  \citenamefont {Dutta}, \citenamefont {Loganayagam},\ and\ \citenamefont
  {Surowka}}]{Banerjee:2008th}%
  \BibitemOpen
  \bibfield  {author} {\bibinfo {author} {\bibfnamefont {N.}~\bibnamefont
  {Banerjee}}, \bibinfo {author} {\bibfnamefont {J.}~\bibnamefont
  {Bhattacharya}}, \bibinfo {author} {\bibfnamefont {S.}~\bibnamefont
  {Bhattacharyya}}, \bibinfo {author} {\bibfnamefont {S.}~\bibnamefont
  {Dutta}}, \bibinfo {author} {\bibfnamefont {R.}~\bibnamefont {Loganayagam}},
  \ and\ \bibinfo {author} {\bibfnamefont {P.}~\bibnamefont {Surowka}},\ }\href
  {\doibase 10.1007/JHEP01(2011)094} {\bibfield  {journal} {\bibinfo  {journal}
  {JHEP}\ }\textbf {\bibinfo {volume} {01}},\ \bibinfo {pages} {094} (\bibinfo
  {year} {2011})},\ \Eprint {http://arxiv.org/abs/0809.2596} {arXiv:0809.2596
  [hep-th]} \BibitemShut {NoStop}%
\bibitem [{\citenamefont {Chesler}\ and\ \citenamefont
  {Yaffe}(2014)}]{Chesler:2013lia}%
  \BibitemOpen
  \bibfield  {author} {\bibinfo {author} {\bibfnamefont {P.~M.}\ \bibnamefont
  {Chesler}}\ and\ \bibinfo {author} {\bibfnamefont {L.~G.}\ \bibnamefont
  {Yaffe}},\ }\href {\doibase 10.1007/JHEP07(2014)086} {\bibfield  {journal}
  {\bibinfo  {journal} {JHEP}\ }\textbf {\bibinfo {volume} {07}},\ \bibinfo
  {pages} {086} (\bibinfo {year} {2014})},\ \Eprint
  {http://arxiv.org/abs/1309.1439} {arXiv:1309.1439 [hep-th]} \BibitemShut
  {NoStop}%
\bibitem [{\citenamefont {Lin}\ and\ \citenamefont {Yee}(2013)}]{Lin:2013sga}%
  \BibitemOpen
  \bibfield  {author} {\bibinfo {author} {\bibfnamefont {S.}~\bibnamefont
  {Lin}}\ and\ \bibinfo {author} {\bibfnamefont {H.-U.}\ \bibnamefont {Yee}},\
  }\href {\doibase 10.1103/PhysRevD.88.025030} {\bibfield  {journal} {\bibinfo
  {journal} {Phys. Rev.}\ }\textbf {\bibinfo {volume} {D88}},\ \bibinfo {pages}
  {025030} (\bibinfo {year} {2013})},\ \Eprint {http://arxiv.org/abs/1305.3949}
  {arXiv:1305.3949 [hep-ph]} \BibitemShut {NoStop}%
\bibitem [{\citenamefont {Ammon}\ \emph {et~al.}(2016)\citenamefont {Ammon},
  \citenamefont {Grieninger}, \citenamefont {Jimenez-Alba}, \citenamefont
  {Macedo},\ and\ \citenamefont {Melgar}}]{Ammon:2016fru}%
  \BibitemOpen
  \bibfield  {author} {\bibinfo {author} {\bibfnamefont {M.}~\bibnamefont
  {Ammon}}, \bibinfo {author} {\bibfnamefont {S.}~\bibnamefont {Grieninger}},
  \bibinfo {author} {\bibfnamefont {A.}~\bibnamefont {Jimenez-Alba}}, \bibinfo
  {author} {\bibfnamefont {R.~P.}\ \bibnamefont {Macedo}}, \ and\ \bibinfo
  {author} {\bibfnamefont {L.}~\bibnamefont {Melgar}},\ }\href {\doibase
  10.1007/JHEP09(2016)131} {\bibfield  {journal} {\bibinfo  {journal} {JHEP}\
  }\textbf {\bibinfo {volume} {09}},\ \bibinfo {pages} {131} (\bibinfo {year}
  {2016})},\ \Eprint {http://arxiv.org/abs/1607.06817} {arXiv:1607.06817
  [hep-th]} \BibitemShut {NoStop}%
\bibitem [{\citenamefont {Landsteiner}\ \emph {et~al.}(2018)\citenamefont
  {Landsteiner}, \citenamefont {Lopez},\ and\ \citenamefont {Milans~del
  Bosch}}]{Landsteiner:2017lwm}%
  \BibitemOpen
  \bibfield  {author} {\bibinfo {author} {\bibfnamefont {K.}~\bibnamefont
  {Landsteiner}}, \bibinfo {author} {\bibfnamefont {E.}~\bibnamefont {Lopez}},
  \ and\ \bibinfo {author} {\bibfnamefont {G.}~\bibnamefont {Milans~del
  Bosch}},\ }\href {\doibase 10.1103/PhysRevLett.120.071602} {\bibfield
  {journal} {\bibinfo  {journal} {Phys. Rev. Lett.}\ }\textbf {\bibinfo
  {volume} {120}},\ \bibinfo {pages} {071602} (\bibinfo {year} {2018})},\
  \Eprint {http://arxiv.org/abs/1709.08384} {arXiv:1709.08384 [hep-th]}
  \BibitemShut {NoStop}%
\bibitem [{\citenamefont {Fernández-Pendás}\ and\ \citenamefont
  {Landsteiner}(2019)}]{Fernandez-Pendas:2019rkh}%
  \BibitemOpen
  \bibfield  {author} {\bibinfo {author} {\bibfnamefont {J.}~\bibnamefont
  {Fernández-Pendás}}\ and\ \bibinfo {author} {\bibfnamefont
  {K.}~\bibnamefont {Landsteiner}},\ }\href {\doibase
  10.1103/PhysRevD.100.126024} {\bibfield  {journal} {\bibinfo  {journal}
  {Phys. Rev. D}\ }\textbf {\bibinfo {volume} {100}},\ \bibinfo {pages}
  {126024} (\bibinfo {year} {2019})},\ \Eprint
  {http://arxiv.org/abs/1907.09962} {arXiv:1907.09962 [hep-th]} \BibitemShut
  {NoStop}%
\bibitem [{\citenamefont {Buividovich}\ and\ \citenamefont
  {Valgushev}(2016)}]{Buividovich:2016ulp}%
  \BibitemOpen
  \bibfield  {author} {\bibinfo {author} {\bibfnamefont {P.}~\bibnamefont
  {Buividovich}}\ and\ \bibinfo {author} {\bibfnamefont {S.}~\bibnamefont
  {Valgushev}},\ }\href {\doibase 10.22323/1.256.0253} {\bibfield  {journal}
  {\bibinfo  {journal} {PoS}\ }\textbf {\bibinfo {volume} {LATTICE2016}},\
  \bibinfo {pages} {253} (\bibinfo {year} {2016})},\ \Eprint
  {http://arxiv.org/abs/1611.05294} {arXiv:1611.05294 [hep-lat]} \BibitemShut
  {NoStop}%
\bibitem [{\citenamefont {Mace}\ \emph {et~al.}(2017)\citenamefont {Mace},
  \citenamefont {Mueller}, \citenamefont {Schlichting},\ and\ \citenamefont
  {Sharma}}]{Mace:2016shq}%
  \BibitemOpen
  \bibfield  {author} {\bibinfo {author} {\bibfnamefont {M.}~\bibnamefont
  {Mace}}, \bibinfo {author} {\bibfnamefont {N.}~\bibnamefont {Mueller}},
  \bibinfo {author} {\bibfnamefont {S.}~\bibnamefont {Schlichting}}, \ and\
  \bibinfo {author} {\bibfnamefont {S.}~\bibnamefont {Sharma}},\ }\href
  {\doibase 10.1103/PhysRevD.95.036023} {\bibfield  {journal} {\bibinfo
  {journal} {Phys. Rev. D}\ }\textbf {\bibinfo {volume} {95}},\ \bibinfo
  {pages} {036023} (\bibinfo {year} {2017})},\ \Eprint
  {http://arxiv.org/abs/1612.02477} {arXiv:1612.02477 [hep-lat]} \BibitemShut
  {NoStop}%
\bibitem [{\citenamefont {Kharzeev}\ and\ \citenamefont
  {Kikuchi}(2020)}]{Kharzeev:2020kgc}%
  \BibitemOpen
  \bibfield  {author} {\bibinfo {author} {\bibfnamefont {D.~E.}\ \bibnamefont
  {Kharzeev}}\ and\ \bibinfo {author} {\bibfnamefont {Y.}~\bibnamefont
  {Kikuchi}},\ }\href@noop {} {\  (\bibinfo {year} {2020})},\ \Eprint
  {http://arxiv.org/abs/2001.00698} {arXiv:2001.00698 [hep-ph]} \BibitemShut
  {NoStop}%
\bibitem [{\citenamefont {Hidaka}\ and\ \citenamefont
  {Yang}(2018)}]{Hidaka:2018ekt}%
  \BibitemOpen
  \bibfield  {author} {\bibinfo {author} {\bibfnamefont {Y.}~\bibnamefont
  {Hidaka}}\ and\ \bibinfo {author} {\bibfnamefont {D.-L.}\ \bibnamefont
  {Yang}},\ }\href {\doibase 10.1103/PhysRevD.98.016012} {\bibfield  {journal}
  {\bibinfo  {journal} {Phys. Rev.}\ }\textbf {\bibinfo {volume} {D98}},\
  \bibinfo {pages} {016012} (\bibinfo {year} {2018})},\ \Eprint
  {http://arxiv.org/abs/1801.08253} {arXiv:1801.08253 [hep-th]} \BibitemShut
  {NoStop}%
\bibitem [{\citenamefont {Landsteiner}\ \emph
  {et~al.}(2011{\natexlab{b}})\citenamefont {Landsteiner}, \citenamefont
  {Megias},\ and\ \citenamefont {Pena-Benitez}}]{Landsteiner:2011cp}%
  \BibitemOpen
  \bibfield  {author} {\bibinfo {author} {\bibfnamefont {K.}~\bibnamefont
  {Landsteiner}}, \bibinfo {author} {\bibfnamefont {E.}~\bibnamefont {Megias}},
  \ and\ \bibinfo {author} {\bibfnamefont {F.}~\bibnamefont {Pena-Benitez}},\
  }\href {\doibase 10.1103/PhysRevLett.107.021601} {\bibfield  {journal}
  {\bibinfo  {journal} {Phys. Rev. Lett.}\ }\textbf {\bibinfo {volume} {107}},\
  \bibinfo {pages} {021601} (\bibinfo {year} {2011}{\natexlab{b}})},\ \Eprint
  {http://arxiv.org/abs/1103.5006} {arXiv:1103.5006 [hep-ph]} \BibitemShut
  {NoStop}%
\bibitem [{\citenamefont {Jensen}\ \emph {et~al.}(2013)\citenamefont {Jensen},
  \citenamefont {Loganayagam},\ and\ \citenamefont {Yarom}}]{Jensen:2012kj}%
  \BibitemOpen
  \bibfield  {author} {\bibinfo {author} {\bibfnamefont {K.}~\bibnamefont
  {Jensen}}, \bibinfo {author} {\bibfnamefont {R.}~\bibnamefont {Loganayagam}},
  \ and\ \bibinfo {author} {\bibfnamefont {A.}~\bibnamefont {Yarom}},\ }\href
  {\doibase 10.1007/JHEP02(2013)088} {\bibfield  {journal} {\bibinfo  {journal}
  {JHEP}\ }\textbf {\bibinfo {volume} {02}},\ \bibinfo {pages} {088} (\bibinfo
  {year} {2013})},\ \Eprint {http://arxiv.org/abs/1207.5824} {arXiv:1207.5824
  [hep-th]} \BibitemShut {NoStop}%
\bibitem [{\citenamefont {Stone}\ and\ \citenamefont
  {Kim}(2018)}]{Stone:2018zel}%
  \BibitemOpen
  \bibfield  {author} {\bibinfo {author} {\bibfnamefont {M.}~\bibnamefont
  {Stone}}\ and\ \bibinfo {author} {\bibfnamefont {J.}~\bibnamefont {Kim}},\
  }\href {\doibase 10.1103/PhysRevD.98.025012} {\bibfield  {journal} {\bibinfo
  {journal} {Phys. Rev.}\ }\textbf {\bibinfo {volume} {D98}},\ \bibinfo {pages}
  {025012} (\bibinfo {year} {2018})},\ \Eprint
  {http://arxiv.org/abs/1804.08668} {arXiv:1804.08668 [cond-mat.mes-hall]}
  \BibitemShut {NoStop}%
\bibitem [{\citenamefont {Golkar}\ and\ \citenamefont
  {Sethi}(2016)}]{Golkar:2015oxw}%
  \BibitemOpen
  \bibfield  {author} {\bibinfo {author} {\bibfnamefont {S.}~\bibnamefont
  {Golkar}}\ and\ \bibinfo {author} {\bibfnamefont {S.}~\bibnamefont {Sethi}},\
  }\href {\doibase 10.1007/JHEP05(2016)105} {\bibfield  {journal} {\bibinfo
  {journal} {JHEP}\ }\textbf {\bibinfo {volume} {05}},\ \bibinfo {pages} {105}
  (\bibinfo {year} {2016})},\ \Eprint {http://arxiv.org/abs/1512.02607}
  {arXiv:1512.02607 [hep-th]} \BibitemShut {NoStop}%
\bibitem [{\citenamefont {Andrade}\ and\ \citenamefont
  {Withers}(2014)}]{Andrade:2013gsa}%
  \BibitemOpen
  \bibfield  {author} {\bibinfo {author} {\bibfnamefont {T.}~\bibnamefont
  {Andrade}}\ and\ \bibinfo {author} {\bibfnamefont {B.}~\bibnamefont
  {Withers}},\ }\href {\doibase 10.1007/JHEP05(2014)101} {\bibfield  {journal}
  {\bibinfo  {journal} {JHEP}\ }\textbf {\bibinfo {volume} {05}},\ \bibinfo
  {pages} {101} (\bibinfo {year} {2014})},\ \Eprint
  {http://arxiv.org/abs/1311.5157} {arXiv:1311.5157 [hep-th]} \BibitemShut
  {NoStop}%
\bibitem [{\citenamefont {Stephanov}\ and\ \citenamefont
  {Yee}(2016)}]{Stephanov:2015roa}%
  \BibitemOpen
  \bibfield  {author} {\bibinfo {author} {\bibfnamefont {M.~A.}\ \bibnamefont
  {Stephanov}}\ and\ \bibinfo {author} {\bibfnamefont {H.-U.}\ \bibnamefont
  {Yee}},\ }\href {\doibase 10.1103/PhysRevLett.116.122302} {\bibfield
  {journal} {\bibinfo  {journal} {Phys. Rev. Lett.}\ }\textbf {\bibinfo
  {volume} {116}},\ \bibinfo {pages} {122302} (\bibinfo {year} {2016})},\
  \Eprint {http://arxiv.org/abs/1508.02396} {arXiv:1508.02396 [hep-th]}
  \BibitemShut {NoStop}%
\bibitem [{\citenamefont {Copetti}\ \emph {et~al.}(2017)\citenamefont
  {Copetti}, \citenamefont {Fernández-Pendás}, \citenamefont {Landsteiner},\
  and\ \citenamefont {Megías}}]{Copetti:2017ywz}%
  \BibitemOpen
  \bibfield  {author} {\bibinfo {author} {\bibfnamefont {C.}~\bibnamefont
  {Copetti}}, \bibinfo {author} {\bibfnamefont {J.}~\bibnamefont
  {Fernández-Pendás}}, \bibinfo {author} {\bibfnamefont {K.}~\bibnamefont
  {Landsteiner}}, \ and\ \bibinfo {author} {\bibfnamefont {E.}~\bibnamefont
  {Megías}},\ }\href {\doibase 10.1007/JHEP09(2017)004} {\bibfield  {journal}
  {\bibinfo  {journal} {JHEP}\ }\textbf {\bibinfo {volume} {09}},\ \bibinfo
  {pages} {004} (\bibinfo {year} {2017})},\ \Eprint
  {http://arxiv.org/abs/1706.05294} {arXiv:1706.05294 [hep-th]} \BibitemShut
  {NoStop}%
\bibitem [{\citenamefont {Alvarez-Gaume}\ and\ \citenamefont
  {Witten}(1984)}]{AlvarezGaume:1983ig}%
  \BibitemOpen
  \bibfield  {author} {\bibinfo {author} {\bibfnamefont {L.}~\bibnamefont
  {Alvarez-Gaume}}\ and\ \bibinfo {author} {\bibfnamefont {E.}~\bibnamefont
  {Witten}},\ }\href {\doibase 10.1016/0550-3213(84)90066-X} {\bibfield
  {journal} {\bibinfo  {journal} {Nucl. Phys.}\ }\textbf {\bibinfo {volume}
  {B234}},\ \bibinfo {pages} {269} (\bibinfo {year} {1984})}\BibitemShut
  {NoStop}%
\bibitem [{\citenamefont {Landsteiner}\ \emph
  {et~al.}(2011{\natexlab{c}})\citenamefont {Landsteiner}, \citenamefont
  {Megias}, \citenamefont {Melgar},\ and\ \citenamefont
  {Pena-Benitez}}]{Landsteiner:2011iq}%
  \BibitemOpen
  \bibfield  {author} {\bibinfo {author} {\bibfnamefont {K.}~\bibnamefont
  {Landsteiner}}, \bibinfo {author} {\bibfnamefont {E.}~\bibnamefont {Megias}},
  \bibinfo {author} {\bibfnamefont {L.}~\bibnamefont {Melgar}}, \ and\ \bibinfo
  {author} {\bibfnamefont {F.}~\bibnamefont {Pena-Benitez}},\ }\href {\doibase
  10.1007/JHEP09(2011)121} {\bibfield  {journal} {\bibinfo  {journal} {JHEP}\
  }\textbf {\bibinfo {volume} {09}},\ \bibinfo {pages} {121} (\bibinfo {year}
  {2011}{\natexlab{c}})},\ \Eprint {http://arxiv.org/abs/1107.0368}
  {arXiv:1107.0368 [hep-th]} \BibitemShut {NoStop}%
\bibitem [{\citenamefont {Blake}\ \emph {et~al.}(2014)\citenamefont {Blake},
  \citenamefont {Tong},\ and\ \citenamefont {Vegh}}]{Blake:2013owa}%
  \BibitemOpen
  \bibfield  {author} {\bibinfo {author} {\bibfnamefont {M.}~\bibnamefont
  {Blake}}, \bibinfo {author} {\bibfnamefont {D.}~\bibnamefont {Tong}}, \ and\
  \bibinfo {author} {\bibfnamefont {D.}~\bibnamefont {Vegh}},\ }\href {\doibase
  10.1103/PhysRevLett.112.071602} {\bibfield  {journal} {\bibinfo  {journal}
  {Phys. Rev. Lett.}\ }\textbf {\bibinfo {volume} {112}},\ \bibinfo {pages}
  {071602} (\bibinfo {year} {2014})},\ \Eprint {http://arxiv.org/abs/1310.3832}
  {arXiv:1310.3832 [hep-th]} \BibitemShut {NoStop}%
\bibitem [{\citenamefont {Zaanen}\ \emph {et~al.}(2015)\citenamefont {Zaanen},
  \citenamefont {Sun}, \citenamefont {Liu},\ and\ \citenamefont
  {Schalm}}]{Zaanen:2015oix}%
  \BibitemOpen
  \bibfield  {author} {\bibinfo {author} {\bibfnamefont {J.}~\bibnamefont
  {Zaanen}}, \bibinfo {author} {\bibfnamefont {Y.-W.}\ \bibnamefont {Sun}},
  \bibinfo {author} {\bibfnamefont {Y.}~\bibnamefont {Liu}}, \ and\ \bibinfo
  {author} {\bibfnamefont {K.}~\bibnamefont {Schalm}},\ }\href
  {http://www.cambridge.org/mw/academic/subjects/physics/condensed-matter-physics-nanoscience-and-mesoscopic-physics/holographic-duality-condensed-matter-physics?format=HB}
  {\emph {\bibinfo {title} {{Holographic Duality in Condensed Matter
  Physics}}}}\ (\bibinfo  {publisher} {Cambridge Univ. Press},\ \bibinfo {year}
  {2015})\BibitemShut {NoStop}%
\bibitem [{\citenamefont {Ammon}\ and\ \citenamefont
  {Erdmenger}(2015)}]{Ammon:2015wua}%
  \BibitemOpen
  \bibfield  {author} {\bibinfo {author} {\bibfnamefont {M.}~\bibnamefont
  {Ammon}}\ and\ \bibinfo {author} {\bibfnamefont {J.}~\bibnamefont
  {Erdmenger}},\ }\href
  {http://www.cambridge.org/de/academic/subjects/physics/theoretical-physics-and-mathematical-physics/gaugegravity-duality-foundations-and-applications}
  {\emph {\bibinfo {title} {{Gauge/gravity duality}}}}\ (\bibinfo  {publisher}
  {Cambridge Univ. Pr.},\ \bibinfo {address} {Cambridge, UK},\ \bibinfo {year}
  {2015})\BibitemShut {NoStop}%
\bibitem [{\citenamefont {Copetti}\ and\ \citenamefont
  {Fernández-Pendás}(2018)}]{Copetti:2017cin}%
  \BibitemOpen
  \bibfield  {author} {\bibinfo {author} {\bibfnamefont {C.}~\bibnamefont
  {Copetti}}\ and\ \bibinfo {author} {\bibfnamefont {J.}~\bibnamefont
  {Fernández-Pendás}},\ }\href {\doibase 10.1007/JHEP04(2018)134} {\bibfield
  {journal} {\bibinfo  {journal} {JHEP}\ }\textbf {\bibinfo {volume} {04}},\
  \bibinfo {pages} {134} (\bibinfo {year} {2018})},\ \Eprint
  {http://arxiv.org/abs/1712.06628} {arXiv:1712.06628 [hep-th]} \BibitemShut
  {NoStop}%
\bibitem [{\citenamefont {Kaminski}\ \emph {et~al.}(2010)\citenamefont
  {Kaminski}, \citenamefont {Landsteiner}, \citenamefont {Mas}, \citenamefont
  {Shock},\ and\ \citenamefont {Tarrio}}]{Kaminski:2009dh}%
  \BibitemOpen
  \bibfield  {author} {\bibinfo {author} {\bibfnamefont {M.}~\bibnamefont
  {Kaminski}}, \bibinfo {author} {\bibfnamefont {K.}~\bibnamefont
  {Landsteiner}}, \bibinfo {author} {\bibfnamefont {J.}~\bibnamefont {Mas}},
  \bibinfo {author} {\bibfnamefont {J.~P.}\ \bibnamefont {Shock}}, \ and\
  \bibinfo {author} {\bibfnamefont {J.}~\bibnamefont {Tarrio}},\ }\href
  {\doibase 10.1007/JHEP02(2010)021} {\bibfield  {journal} {\bibinfo  {journal}
  {JHEP}\ }\textbf {\bibinfo {volume} {02}},\ \bibinfo {pages} {021} (\bibinfo
  {year} {2010})},\ \Eprint {http://arxiv.org/abs/0911.3610} {arXiv:0911.3610
  [hep-th]} \BibitemShut {NoStop}%
\end{thebibliography}%

\end{document}